\documentclass[final,5p,10pc,times,twocolumn]{elsarticle}

\usepackage{lineno}
\modulolinenumbers[5]
\usepackage{scalerel}
\usepackage{slantsc}
\usepackage{bm}
\usepackage{dsfont}
\usepackage{ifpdf}
\ifpdf
\usepackage[%
pdftitle={Covariance-based anisotropic smoothed particle hydrodynamics},%
pdfauthor={Eraldo P Marinho},%
pdfsubject={The preprint document class elsart},%
pdfkeywords={instructions for use, elsart, document class},%
pdfstartview=FitH,%
bookmarks=true,%
bookmarksopen=true,%
breaklinks=true,%
colorlinks=true,%
linkcolor=blue,anchorcolor=blue,%
urlcolor=blue]{hyperref}
\else
\usepackage[%
breaklinks=true,%
colorlinks=true,%
linkcolor=blue,anchorcolor=blue,%
citecolor=blue,filecolor=blue,%
menucolor=blue,pagecolor=blue,%
urlcolor=blue]{hyperref}
\fi
\usepackage{graphics}
\usepackage{graphicx}
\usepackage{algorithm}
\usepackage{algpseudocode}
\usepackage{longtable}
\usepackage{amsfonts}
\usepackage{amssymb}
\usepackage{amsmath}
\usepackage{theorem}
\usepackage{media9}
\renewcommand{\vec}[1]{\bm{#1}}
\newcommand{\set}[1]{\{#1\}}
\newcommand{\bigset}[1]{\bigl\{#1\bigr\}}

\newcommand{\tensor}[1]{\mathbf{#1}}
\newcommand{\trace}{\text{tr\ }}
\newcommand{\vecspc}[2][3]{\mathbb{#2}^{#1}}
\newcommand{\deriv}[2][t]{\frac{\text{d}#2}{\text{d}#1}}
\newcommand{\pderiv}[2][t]{\frac{\partial#2}{\partial#1}}
\newcommand{\transpose}[1]{{#1}^t}
\journal{Journal of Computational Physics}

\biboptions{authoryear}

\makeindex

\begin{document}

\begin{frontmatter}

\title{Covariance-based smoothed particle hydrodynamics. A 
machine-learning application to simulating disc fragmentation}
\author{Eraldo Pereira Marinho}
\address{Sao Paulo State University (UNESP),
Department of Statistics, Applied Mathematics and Computing -
Avenida 24A 1515, Rio Claro, Sao Paulo, Brazil}
\ead{pereira.marinho@unesp.br}




\begin{abstract}
A PCA-based, machine learning version of the SPH method is proposed. In the 
present scheme, the smoothing tensor is computed to have their eigenvalues 
proportional to the covariance's principal components, using a modified octree 
data structure, which allows the fast estimation of the anisotropic 
self-regulating kNN. Each SPH particle is the center of such an optimal kNN 
cluster, i.e., the one whose covariance tensor allows the find of the kNN 
cluster itself according to the Mahalanobis metric. Such machine learning 
constitutes a fixed point problem. The definitive (self-regulating) kNN cluster 
defines the smoothing volume, or properly saying, the smoothing ellipsoid, 
required to perform the anisotropic interpolation. Thus, the smoothing kernel 
has an ellipsoidal profile, which changes how the kernel gradients are 
computed. As an application it was performed the simulation of collapse 
and fragmentation of a non-magnetic, rotating gaseous sphere. An interesting 
outcome was the formation of protostars in the disc fragmentation, shown to be 
much more persistent and much more abundant in the anisotropic simulation than 
in the isotropic case.
\end{abstract}
\begin{keyword}
anisotropic density estimation\sep adaptive smoothed particle hydrodynamics\sep
k-nearest neighbor search\sep PCA-based machine learning
\end{keyword}

\end{frontmatter}


\section{Introduction}

The present work consists of combining machine learning with the numerical 
methods of smoothed particle hydrodynamics (SPH). The machine learning approach 
is used to find the optimal smoothing volumes that best express the anisotropic 
tendencies of the particle distribution, as previously proposed by 
\cite{Marinho:2014}. These ideal volumes are defined in terms of the 
self-regulating kNN clusters, as will be seen later.

SPH is a Lagrangian computational method of fluid mechanics that was introduced 
by two independent works \citep{GingoldMonaghan:1977,Lucy:1977}. On the other 
hand, machine learning consists of an algorithm, or a combination of 
algorithms, that automatically refines the approximate solution to a pattern 
recognition problem, allowing future classifications to be more efficient once 
the machine has acquired experience from the previous classification 
\citep{DudaHartStork:2001}.


One of the main problems of SPH concerns the best morphology of the smoothing
volume, regarding the optimal spatial resolution. A first attempt to
improve spatial resolution is to consider each simulation particle (say, query
particle) as the center of its $k$-nearest neighbors 
\citep[e.g.][]{Monaghan:1992,Monaghan:1994,Monaghan:2012}, which in turn is a 
variant of the Parzen window \citep[e.g.][]{DudaHartStork:2001}. In such 
estimation technique the smoothing kernel is spherical, with smoothing length 
proportional to $\rho^{-1/3}$. Therefore, the orthodox adaptive SPH is in fact a
density-adaptive scheme, which in principle disregard the multivariate
distribution of the simulation particles.

A second advance made in order to obtain a more complete spatial adaptability,
both in density and in the preferential direction of deformation of the
smoothing volume, was proposed by 
\cite[][hereafter MSVK]{MartelShapiroVillumsenKang:1993}, which named the 
adaptive technique as ASPH, followed by the works of 
\cite{ShapiroMartelVillumsenOwen:1996}, \cite{OwenVillumsenShapiroMartel:1998} 
and \cite{MartelShapiro:2003}. The authors introduced a tensor version of the 
smoothing length, namely, the smoothing tensor $\tensor{H}$, whose components 
change in space and time according to the estimated deformation-rate tensor, 
$\vec\nabla\vec{v}$. The squared smoothing tensor $\tensor{H}^2$ is the metric 
tensor whose minimal surface of equidistant points containing all the kNN 
corresponds to the ellipsoid whose semi-major axes are the eigenvalues 
(eigenvectors) of $\tensor{H}$.

In the present work, a machine learning technique is proposed, which in a
way resembles the MSVK technique, but is based on the principal components
analysis (PCA), using the anisotropic search method for the $k$-nearest
neighbors ($k$-NN) proposed by \cite{Marinho-Andreazza:2010}. The
Marinho--Andreazza approach is an unsupervised machine learning method of
finding the optimal $k$-NN set, under the Mahalanobis metric
\citep{Mahalanobis:1936}, named here as the self-regulating $k$-NN cluster. The
proposed anisotropic SPH code is a covariance-based SPH, addressed hereafter as
Sigma-SPH.

What is new in this work is that the anisotropic kernel is defined only by
multivariate arguments, involving the properties of the covariance tensor,
without taking into account the ASPH technique. In summary, Sigma-SPH is based
solely on the multivariability of the particle distribution, while in ASPH the
smoothing tensor is defined according to the dynamics of the simulated fluid,
via deformation rate tensor.

Presently, the smoothing tensor is proportional to the squared root of the 
covariance tensor, namely $\tensor{H}=\gamma_\mathrm{max}\tensor\Sigma^{1/2}$, 
where $\gamma_\mathrm{max}$ is the Mahalanobis distance from the query particle 
to the outermost one in the self-regulating $k$-NN cluster. Thus, $\tensor{H}$ 
is the tensor whose eigenvalues/eigenvectors sets up the minimal ellipsoid hull 
of the self-regulating $k$-NN cluster.


To continue, we will use pattern recognition terminology, such as the dataset 
concept, adapted for the set of particles used in the SPH simulation. An SPH 
dataset is seen as a collection of classic $N$ particles, seen as massive 
points in the configuration space. Each particle has then three coordinates 
for positions and three coordinates for velocities. It is assumed that the 
particles are distinguishable only by their positions, regardless of their other 
physical attributes. The coordinates of the particles make up a unique query 
key. Further relevant quantities are attributed to the particles in the 
dataset, such as specific thermal energy. Of course, each particle is 
represented by an instance in the dataset.

In the present work, we are concerned with the three-dimensional description of
SPH particles in euclidean space $\vecspc{E}$ so that velocities and other
intrinsic quantities are field functions of spatial coordinates. Time is 
naturally implicit in the Lagrangian equations of fluid motion.

For simplicity and mathematical conciseness, we abstract the dataset as a
collection of identifiers or indexes, which form a bijection with the spatial
coordinates of the particles. Thus, we refer to the dataset as simply the
index collection $\mathcal{D} = \{1,2, \ldots, N\}$ whenever necessary. Any 
subset $\mathcal{C}$ of
$\mathcal{D}$ is called a cluster. In particular, we are primarily interested 
in 
clusters
that constitute a partition of $\mathcal{D}$. For instance, $\mathcal{A}$ and
$\mathcal{B}$ constitute a cluster partition if and only if
$\mathcal{A}\cup\mathcal{B}=\mathcal{D}$ and
$\mathcal{A}\cap\mathcal{B}=\emptyset$.

If $\mathcal{C}$ is a cluster within $\mathcal{D}$, we assume the following
correspondence:
\begin{equation}
 \mathcal{C}=\{i_1,i_2,\ldots,i_k\}\subseteq\{1,\ldots,N\}
\end{equation}
Thus, the cluster's total mass is written as
\begin{equation}\label{eq:02}
 M_\mathcal{C}=\sum_{j\in\mathcal{C}}m_j
\end{equation}
The cluster's center of mass, aka cluster's mean position, is written as the
following normalized first momentum:
\begin{equation}\label{eq:03}
 \vec{r}_\mathcal{C}=\frac1{M_\mathcal{C}}\sum_{j\in\mathcal{C}}m_j\vec{r}_j
\end{equation}
The cluster's covariance tensor \citep[e.g.][]{DudaHartStork:2001} is defined as
\begin{equation}
 \tensor{\Sigma}_\mathcal{C}=\frac1{M_\mathcal{C}}\sum_{j\in\mathcal{C}}
    m_j(\vec{r}_j-\vec{r}_\mathcal{C})\otimes(\vec{r}_j-\vec{r}_\mathcal{C})
\end{equation}
$\otimes$ stands for the tensor product, defined here as the outer product
$\otimes:\vecspc{E}\times\vecspc{E}\mapsto\vecspc[3^2]{E}$, which in matrix
notation corresponds to
\begin{equation}\label{eq:05}
 \vec A\otimes \vec B\equiv \vec A\vec B^t
\end{equation}
where
\begin{equation}\label{eq:06}
 \vec A\equiv\begin{pmatrix}
         A_1\\
         A_2\\
         A_3
        \end{pmatrix}\in\vecspc{R}
        \;\text{and}\;
 \vec B^t\equiv\begin{pmatrix}
                B_1 B_2 B_3
               \end{pmatrix}\in(\vecspc{R})^t
\end{equation}
and $(\vecspc{R})^t$ is the dual space of $\vecspc{R}$.
Thus, from equation~(\ref{eq:05}),
\begin{equation}\label{eq:07}
 \vec A\otimes\vec B\equiv \vec A\vec B^t\equiv\begin{pmatrix}
                      A_1B_1&A_1B_2&A_1B_3\\
                      A_2B_1&A_2B_2&A_2B_3\\
                      A_3B_1&A_3B_2&A_3B_3
                     \end{pmatrix}\in\vecspc{R}\otimes\vecspc{R}
\end{equation}
where the tensor-product space, $\vecspc{R}\otimes\vecspc{R}$, is isomorphic to 
the vector space of the real $3\times3$ square matrices, namely 
$\vecspc{R}\otimes\vecspc{R}\equiv\vecspc[3^2]{R}$.

On the other hand, the scalar product $\vec A\cdot\vec B$ has its matrix 
representation given by
\begin{equation}\label{eq:08}
 \vec A\cdot\vec B\equiv \vec A^t\vec B
\end{equation}
It follows immediately from the definitions for $\vec A$ and $\vec B$, 
previously given in equation~(\ref{eq:06}), that
\begin{equation}\label{eq:09}
 \vec A\cdot\vec B\equiv\vec A^t\vec B\equiv A_1B_1+A_2B_2+A_3B_3=\trace\vec 
A\vec B^t
\end{equation}
where $\trace\vec{A}\transpose{\vec{B}}$ is the trace of the matrix 
$\vec{A}\transpose{\vec{B}}$ according to equations~(\ref{eq:05}) and 
(\ref{eq:07})

For notation brevity, and regarding equation~(\ref{eq:05}), the covariance 
tensor can be written simply as
\begin{equation}\label{eq:10}
 \tensor{\Sigma}_\mathcal{C}=\frac1{M_\mathcal{C}}\sum_{j\in\mathcal{C}}
    m_j(\vec{r}_j-\vec{r}_\mathcal{C})(\vec{r}_j-\vec{r}_\mathcal{C})^t
\end{equation}

The cluster's variance $\sigma_\mathcal{C}^2$ is computed as
\begin{equation}\label{eq:11}
 {\sigma}_\mathcal{C}^2=\frac1{M_\mathcal{C}}\sum_{j\in\mathcal{C}}
    m_j(\vec{r}_j-\vec{r}_\mathcal{C})^t(\vec{r}_j-\vec{r}_\mathcal{C})
    \equiv\trace\tensor{\Sigma}_\mathcal{C}
\end{equation}
which is the isotropic measure of the cluster's dispersion, in contrast to the
fact that $\tensor\Sigma_\mathcal{C}$ measures the anisotropic data dispersion.
The latter result, 
${\sigma}_\mathcal{C}^2\equiv\trace\tensor{\Sigma}_\mathcal{C},$
comes from equation~(\ref{eq:09}).

Since $\tensor\Sigma_\mathcal{C}$ is positive definite, it comes immediately 
that $$\trace\tensor{\Sigma}_\mathcal{C}>0.$$
The ${\sigma}_\mathcal{C}^2=0$ case is not considered since the
entire dataset $\mathcal{D}$ would degenerate into a single particle. 

One cautionary remark is that the number of non-zero eigenvalues of 
the tensor $\tensor{\Sigma}_\mathcal{C}$ depends on the distribution topology. 
For instance, if the entire dataset degenerates into a plane surface, it means 
that there is one null eigenvalue. Such a situation is almost improbable in a 
3D SPH simulation due to the initial conditions' randomization or, maybe, to the 
3D crystalline initial configuration, combined with the freedom of motion an SPH
particle has; see, for instance, the later section on the SPH equations of
motion. A difficulty can happen in situations of strong compressive shocks, 
which can reduce the shock thickness to almost zero. In this case, some 
tolerance artifice must be used to prevent singularities.

The vector collection 
\begin{equation}
\vec p=
\set{
    \sigma_1\vec{e}_1,\;
    \sigma_2\vec{e}_2,\;
    \sigma_3\vec{e}_3
    }
\end{equation}
is named the principal components of the covariance tensor 
$\tensor\Sigma_\mathcal{C}$, whereas the ordered set
\begin{equation}
\vec\sigma=
\set{\sigma_1^2,\, \sigma_2^2,\, \sigma_3^2
\;\vert\; 
\sigma_1\le\sigma_2\le\sigma_3}
\end{equation}
is the collection of eigenvalues of $\tensor\Sigma_\mathcal{C}$. The collection 
$\{\vec{e}_1, \vec{e}_2, \vec{e}_3\}$ of unit vectors corresponds to the 
normalized eigenvectors of the tensor $\tensor\Sigma_\mathcal{C}$.
Consequently, the diagonal representation of the covariance tensor, adopting 
the convention~(\ref{eq:05}), is given by
\begin{equation}\label{eq:12}
 \tensor\Sigma_\mathcal{C}=
 \sigma_1^2\,\vec{e}_1\vec{e}_1^t +
 \sigma_2^2\,\vec{e}_2\vec{e}_2^t +
 \sigma_3^2\,\vec{e}_3\vec{e}_3^t
\end{equation}

It follows straightforwardly that the diagonal form of the inverse covariance
tensor can be written as in an analogous form of equation~(\ref{eq:12}):
\begin{equation}\label{eq:13}
 \tensor\Sigma_\mathcal{C}^{-1}=
 \frac1{\sigma_1^2}\,\vec{e}_1\vec{e}_1^t +
 \frac1{\sigma_2^2}\,\vec{e}_2\vec{e}_2^t +
 \frac1{\sigma_3^2}\,\vec{e}_3\vec{e}_3^t
\end{equation}

The Mahalanobis distance, which is a measure of how a data point varies toward
different directions about the mean cluster position, $\vec{r}_\mathcal{C}$, is
defined as
\begin{equation}\label{eq:14}
 \delta^2=(\vec{r}-\vec{r}_\mathcal{C})\cdot
 \tensor\Sigma_\mathcal{C}^{-1}\cdot
 (\vec{r}-\vec{r}_\mathcal{C})
 \equiv
 (\vec{r}-\vec{r}_\mathcal{C})^t
 \tensor\Sigma_\mathcal{C}^{-1}
 (\vec{r}-\vec{r}_\mathcal{C})
\end{equation}

Observing equation~(\ref{eq:13}), and assuming that the eigenvalues and
eigenvectors are both known, we have a very simplified form of
the latter equation:
\begin{equation}\label{eq:15}
\delta^2=
  \biggl[\frac{(\vec{r}-\vec{r}_\mathcal{C})\cdot\vec{e}_1}{\sigma_1}\biggr]^2
  +
  \biggl[\frac{(\vec{r}-\vec{r}_\mathcal{C})\cdot\vec{e}_2}{\sigma_2}\biggr]^2
  +
  \biggl[\frac{(\vec{r}-\vec{r}_\mathcal{C})\cdot\vec{e}_3}{\sigma_3}\biggr]^2
\end{equation}

The collection of points in $\vecspc{E}$ whose Mahalanobis distance 
$\delta(\vec{r})$ to the origin equals the unit is defined as the confidence 
ellipsoid, namely,
\begin{equation}\label{eq:16}
 \mathcal{E}_\mathcal{C}^3=\{\vec{r}\in\vecspc{E}\vert
 \delta(\vec{r})=1\}
\end{equation}
which is equivalent to the solution for the equation
\begin{equation}\label{eq:17}
   \biggl[\frac{(\vec{r}-\vec{r}_\mathcal{C})\cdot\vec{e}_1}{\sigma_1}\biggr]^2
  +
  \biggl[\frac{(\vec{r}-\vec{r}_\mathcal{C})\cdot\vec{e}_2}{\sigma_2}\biggr]^2
  +
  \biggl[\frac{(\vec{r}-\vec{r}_\mathcal{C})\cdot\vec{e}_3}{\sigma_3}\biggr]^2
  =1
\end{equation}
Particularly, doing the following transform
\begin{equation}\label{eq:18}
\xi_1=\frac{(\vec{r}-\vec{r}_\mathcal{C})\cdot\vec{e}_1}{\sigma_1},\;
\xi_2=\frac{(\vec{r}-\vec{r}_\mathcal{C})\cdot\vec{e}_2}{\sigma_2},\;
\xi_3=\frac{(\vec{r}-\vec{r}_\mathcal{C})\cdot\vec{e}_3}{\sigma_3},
\end{equation}
we have
\begin{equation}\label{eq:19}
 \xi_1^2+\xi_2^2+\xi_3^2=1,
\end{equation}
which is the equation of a unit sphere $\mathcal{S}_1$ in the uncorrelated 
vector space 
\begin{equation}\label{eq:26}
\vecspc{U}_\mathcal{C}=
\set{
    \vec\xi\;\vert\;\;\vec\xi=\tensor{\Sigma}_\mathcal{C}^{-1/2}
    (\vec{r}-\vec{r}_\mathcal{C})
}
\end{equation}
whose position vector $\vec\xi$ have coordinates 
$(\xi_1,\xi_2,\xi_3)$. Such a vector space, spanning at the average cluster's 
position $\vec{r}_\mathcal{C},$ in which the Mahalanobis metric is transformed 
into the euclidean one, is named the uncorrelated vector space. 
Thus, it is given any vector $\vec{r}\in\vecspc{E}$ an uncorrelated vector 
$\vec\xi\in\vecspc{U}_\mathcal{C}$, once known the mean position 
$\vec{r}_\mathcal{C}$, according to the normalization given in (\ref{eq:18}), 
namely
\begin{equation}\label{eq:20}
 \vec\xi=\xi_1\vec{e}_1+\xi_2\vec{e}_2+\xi_3\vec{e}_3
\end{equation}
It has been used in equation~(\ref{eq:26}) the square root of the 
covariance tensor in its diagonal form, namely,
\begin{equation}
 \tensor\Sigma_\mathcal{C}^{1/2}\equiv
 \sigma_1\vec{e}_1\vec{e}_1^t+
 \sigma_2\vec{e}_2\vec{e}_2^t+
 \sigma_3\vec{e}_3\vec{e}_3^t,
\end{equation}
whose inverse $\tensor\Sigma_\mathcal{C}^{-1/2}$ can be easily computed as
\begin{equation}\label{eq:22}
 \tensor\Sigma_\mathcal{C}^{-1/2}\equiv
 \frac{\vec{e}_1\vec{e}_1^t}{\sigma_1}+
 \frac{\vec{e}_2\vec{e}_2^t}{\sigma_2}+
 \frac{\vec{e}_3\vec{e}_3^t}{\sigma_3}.
\end{equation}


One shall notice that the square module of $\vec\xi\in\vecspc{U}_\mathcal{C}$ 
is computed accordingly to the following dot-product:
\begin{equation}\label{eq:24}
 |\vec\xi|^2=\tensor{\Sigma}_\mathcal{C}^{-1/2}(\vec{r}-\vec{r}_\mathcal{C})
 \cdot\tensor{\Sigma}_\mathcal{C}^{-1/2}(\vec{r}-\vec{r}_\mathcal{C})
 \equiv
 (\vec{r}-\vec{r}_\mathcal{C})^t\tensor{\Sigma}_\mathcal{C}^{-1}(\vec{r}-\vec
r_\mathcal{C}),
\end{equation}
which turns back to the Mahalanobis distance from $\vec{r}$ to the mean 
$\vec{r}_\mathcal{C}$ under the covariance tensor 
$\tensor{\Sigma}_\mathcal{C}$. On the other hand, one finds from 
equations~(\ref{eq:18}) and (\ref{eq:20}) that
\begin{equation}
 |\vec\xi|=\sqrt{\xi_1^2+\xi_2^2+\xi_3^2}
\end{equation}
is the euclidean distance from $\vec\xi$ to the origin of the uncorrelated 
space $\vecspc{U}_\mathcal{C}$. Thus, equation~(\ref{eq:19}) connects the 
confidence ellipsoid $\mathcal{E}_\mathcal{C}^3$ in $\vecspc{E}$ with the unit 
sphere 
$\mathcal{S}_\mathcal{C}^3=
\set{\vec\xi\in\vecspc{U}_\mathcal{C}\;\vert\;\;|\vec\xi|=1}.$

The original space $\vecspc{E}$ from which the dataset points have their
spatial coordinates is called hereafter the correlated space once their points
are correlated according to the covariance tensor. Thus, from the 
equation~(\ref{eq:26}), we have the affine correlated vector space
\begin{equation}\label{eq:27}
\vecspc{E}=
\set{
    \vec{r}\;\vert\;\vec{r}=\tensor{\Sigma}_\mathcal{C}^{1/2}
    \vec\xi+\vec{r}_\mathcal{C}
}.
\end{equation}

transformed

We can generalize the Mahalanobis distance so that it is no longer restricted 
to its statistical meaning. Thus, if $\vec{a}$ and $\vec{b}$ are vectors in 
$\vecspc{E}$, their quadratic Mahalanobis distance is defined as follows
\begin{equation*}
 \delta^2(\vec a, \vec b)=
 (\vec a-\vec b)\cdot
 \tensor\Sigma_\mathcal{C}^{-1}\cdot
 (\vec a-\vec b)
\end{equation*}
\begin{equation}\label{eq:28}
 \equiv
 (\vec a-\vec b)^t
 \tensor\Sigma_\mathcal{C}^{-1}
 (\vec a-\vec b).
\end{equation}

Transforming both $\vec{a}$ and $\vec{b}$ of $\vecspc{E}$ into the vectors 
$\vec\xi_a$ and $\vec\xi_b$ of the $\vecspc{U}_\mathcal{C}$ by means of 
definition~(\ref{eq:26}), respectively, one can easily see that 
equation~(\ref{eq:28}) is equivalent to the euclidean distance in 
$\vecspc{U}_\mathcal{C}$:
\begin{equation}
 \delta^2(\vec{a},\vec{b})\equiv|\vec\xi_a-\vec\xi_b|^2
\end{equation}
Moreover, replacing $\tensor\Sigma_\mathcal{C}$ in equation~(\ref{eq:28}) with 
the identity tensor $\tensor{1}$, we have the quadratic form of the euclidean 
distance, namely,
\begin{equation}\label{eq:30}
 \delta^2(\vec a, \vec b)=
 (\vec a-\vec b)\cdot
 \tensor1\cdot
 (\vec a-\vec b)=|\vec a-\vec b|^2.
\end{equation}
The latter result will be useful in initializing the recursive self-regulating 
kNN.


The purpose of the present work is to present and validate a computer program 
based on machine learning to perform fully adaptive SPH simulations, i.e., 
adaptive to the anisotropic nature of mass distribution in critical situations 
of shock and filamentary fragmentation. The paper is structured as follows.


In Sec.~\ref{sec:02} is discussed the self-regulating kNN cluster machine 
learning approach. In Sec.~\ref{sec:03} it is shown how to estimate the 
ellipsoidal hull for the self-regulating kNN cluster and consequently computing 
the smoothing tensor. The anisotropic model for the smoothing kernel is proposed 
in Sec.~\ref{sec:04}. The anisotropic smoothed particle hydrodynamics are 
discussed in Sec.~\ref{sec:05}. The anisotropic artificial viscosity model is 
proposed in Sec.~\ref{sec:06}. A brief description of the covariance-octree 
based gravity estimation as a modification to the \cite{BarnesHut:1986} method 
is made in Sec.~\ref{sec:07}. The adaptive multiple time-scale leapfrog is 
discussed in Sec.~\ref{sec:08}. As an application, it was performed a simulation 
of the collapse of a rotating gas sphere, which converges to a protostellar like 
disc, which will be discussed in details in Sec.~\ref{sec:10}. Discussion and 
conclusion are made in Sec.~\ref{sec:11}.

\section{Self-regulating kNN cluster}\label{sec:02}

This section presents a machine learning approach to find the anisotropic 
self-regulating kNN. The term self-regulating comes from the fact that the 
covariance tensor estimated over such an ideal cluster is the same tensor used 
to search back for the same kNN cluster according to the Mahalanobis metric. 
Thus, the method is a kind of fixed-point problem. If the learned k-NN cluster 
is reorganized so that its particles displace by a small amount, then the 
computational effort to find the new self-regulating cluster is small compared 
to the initial training. So we say that the method learns how the particles are 
distributed in a multivariate way.

The anisotropic kNN method is an approach for searching for the k-nearest 
neighbors of a query point accordingly to some tensor metric, introduced by 
\cite{Marinho-Andreazza:2010}. However, in that paper, the authors focused more 
on the proposed data structure: the covariance (hyper) quadtree, which allows 
the automatic reduction of dimensionality, proper of the PCA technique. This 
definition of anisotropic kNN is being rescued here in the form of an 
application to effectively determine the smoothing kernel's compact support, 
which will be adopted later in the presently proposed anisotropic SPH. As the 
current purpose is to perform three-dimensional simulations, we renamed the data 
structure of Marinho--Andreazza as the covariance octree instead of the 
covariance quadtree. The self-regulating kNN cluster method is depicted as 
follows.

It is presumed that we already have an anisotropic kNN function as the method 
prescribed by \cite{Marinho-Andreazza:2010}, namely,
\begin{equation}\label{eq:31}
 \mathcal{N}(q_0)=\set{q_0, q_1, \ldots, q_k},
\end{equation}
given the index $q_0$ of the query particle located at position 
$\vec{r_{q_0}}$, and a predicted covariance tensor $\tensor\Sigma$ to perform 
the search according to the Mahalanobis metric.

The positive closure $\mathcal{N}^+$ of the anisotropic kNN is given buy the 
subset $\mathcal{N}^+(q_0)=\mathcal{N}(q_0)-\set{q_0}
=\set{q_1, \ldots, q_k}$, which is the index set of the first $k$ nearest 
proper neighbors from $q_0$. Moreover, $\mathcal{N}(q_0)$ is 
ordered by Mahalanobis distances, namely, $\delta(\vec{r}_{q_1},\vec{r}_{q_0}) 
\le \delta(\vec{r}_{q_2},\vec{r}_{q_0}) \le \cdots \le 
\delta(\vec{r}_{q_k},\vec{r}_{q_0})$. If the particles are randomly
distributed, the equality would very difficultly occur. Of course, $\delta(\vec
r_{q_0}, \vec{r}_{q_0}) = 0$ is a trivial result, and, in this case, $q_0$ is 
called an improper neighbor to itself. On the other hand, if $\delta(\vec
r_{p}, \vec{r}_{q}) \ne 0$, then $p$ is said to be a proper neighbor of $q$. 
The set $\mathcal{N}(q_0)$ of the anisotropic kNN constitutes the set of 
training points for the proposed machine learning method.

The initial step of the first training is made by using the identity tensor 
$\tensor1$ in place of the predicted covariance tensor $\tensor\Sigma$. 
By first training we mean the start approach in the beginning of the SPH 
simulation, before performing the integration scheme. Of course, such 
initialization switches the distance from anisotropic to euclidean. Thus, it 
gets the first attempt 
$\mathcal{C}^{(1)}_k\equiv\mathcal{N}(q_0|\tensor1)=\set{q_0,q_1,\ldots,q_k}$, 
which is of course isotropic. Thus, the $\mathcal{C}^{(1)}_k$ morphology 
is almost spherical if it is within $\mathcal{D}$ and far from its borders.

The next approach consists of iteratively computing the covariance tensor for 
the newly found anisotropic neighborhood to predict a neighborhood closer to the
objective that is the self-regulating kNN cluster. Thus, as it has been 
computed $\mathcal{C}^{(n)}_k$ at iteration $n$, compute the covariance tensor 
$\tensor\Sigma_{\mathcal{C}_k}^{(n)}$ for the newly found kNN cluster, and then 
estimate an ever more refined kNN list, namely
\begin{equation}
\mathcal{C}^{(n+1)}_k\gets
\mathcal{N}(q_0|\tensor\Sigma_{\mathcal{C}_k}^{(n)}).
\end{equation}

To recall, the covariance tensor is computed accordingly to 
equation~(\ref{eq:10}) as follows
\begin{equation}\label{eq:32}
 \tensor\Sigma_{\mathcal{C}_k}=
 \frac1{M_{\mathcal{C}_k}}
 \sum_{j\in\mathcal{C}_k}
 m_j(\vec{r}_j-\vec{r}_{\mathcal{C}_k})(\vec{r}_j-\vec{r}_{\mathcal{C}_k})^t
\end{equation}
where $\mathcal{C}_k$ is the anisotropic kNN cluster, having center of mass 
$\vec{r}_{\mathcal{C}_k}$. 
The quantities $M_{\mathcal{C}_k}$ and $\vec{r}_{\mathcal{C}_k}$ are computed 
from equations~(\ref{eq:02})
and (\ref{eq:03}), respectively.

In order to perform the anisotropic kNN search, it is necessary to first
estimate the eigenvalues/eigenvectors of the predicted 
$\tensor\Sigma_{\mathcal{C}_k}$ to have the right-hand side of 
equation~(\ref{eq:22}). It was presently adopted the power-iteration method to 
find a good approximation for eigenvalues. The method usually takes less than 10 
steps to estimate both the eigenvalues and eigenvectors, and a similar count 
takes to have convergence to the self-regulating kNN per query. The power 
iteration method in the present code is just a heritage of the former code to 
perform cluster analysis and PCA in higher dimension spaces 
\citep[e.g.,][]{Marinho-Andreazza:2010}.

According to the steps described above, it can be seen that as the method 
approaches convergence, a central part of the kNN cluster remains unchanged, 
leaving only a tenuous outer margin that is not yet conclusive. Certainly, this 
small population corresponds to particles outside the confidence ellipsoid. This 
residual population has only a few particles compared to the predefined number 
of nearest neighbors, presumably large enough, say, $k\gg1$. However, this 
reasoning only works if the kNN cluster is far enough from the edges of the 
distribution.

Consistent with what was said in the previous paragraph, since the central part 
of the outgoing kNN cluster remains unchanged, it is immediate that the query's 
position is increasingly closer to the center of mass of the kNN cluster. The 
empirical results show that convergence occurs, and the self-regulating kNN 
cluster has its center of mass located very approximates to the query position.

From equations~(\ref{eq:31}) and (\ref{eq:32}), we find
\begin{equation}
 \tensor\Sigma_{\mathcal{C}_k}=
 \frac1{M_{\mathcal{C}_k}}
 \sum_{j=1}^k
    m_{q_j}
    (\vec{r}_{q_j}-\vec{r}_{\mathcal{C}_k})
    (\vec{r}_{q_j}-\vec{r}_{\mathcal{C}_k})^t
    +
    \Delta\tensor\Sigma_{\mathcal{C}_k}
\end{equation}
where the residual covariance tensor is written as
\begin{equation}
\Delta\tensor\Sigma_{\mathcal{C}_k}=
    \frac{m_{q_0}}{M_{\mathcal{C}_k}}
    (\vec{r}_{q_0}-\vec{r}_{\mathcal{C}_k})
    (\vec{r}_{q_0}-\vec{r}_{\mathcal{C}_k})^t
\end{equation}

In the stable configuration, when the residual covariance tensor goes to zero, 
$\Delta\tensor\Sigma_{\mathcal{C}_k}\to\tensor0$, as the query position 
approaches the cluster's center of mass, 
$\vec{r}_{q_0}\to\vec{r}_{\mathcal{C}_k}$, the output kNN cluster becomes 
self-regulating, and then we have the following pair of self-consistent 
equations:
\begin{equation}\label{eq:35}
 \tensor\Sigma_{\mathcal{C}_k^\text{sr}}=
 \frac1{M_{\mathcal{C}_k^\text{sr}}}
 \sum_{j=1}^k
    m_{q_j}
    (\vec{r}_{q_j}-\vec{r}_{q_0})
    (\vec{r}_{q_j}-\vec{r}_{q_0})^t,
\end{equation}
and
\begin{equation}\label{eq:36}
\mathcal{C}_k^\text{sr}=
\mathcal{N}(q_0|\tensor\Sigma_{\mathcal{C}_k^\text{sr}}).
\end{equation}

Repeating what was said in paragraphs before, the reasoning above is only valid 
if the query $q_0$ is within $\mathcal{D}$'s distribution and far from its 
boundary. Considering a particle to be eccentric about its own $k$-nearest 
neighbors may depend on the required number $k$ of nearest neighbors. For 
example, if $k$ were small, it could be that the particle was approximately in 
the middle of its vicinity. On the other hand, increasing $k$ could increase 
the eccentricity of the particle in relation to the center of its neighborhood 
if the particle is near the edges of the entire distribution. The problem of 
the query particle being away from the distribution can be treated as follows.

Suppose the query particle is out of the distribution. In this case, it is 
impossible to have the query as the center of the kNN cluster, at least for a 
preset $ k $ versus the distribution morphology. The algorithm must choose the 
kNN configuration with the center of the cluster as close as possible to the 
query position. Therefore, equations~(\ref{eq:35}) and (\ref{eq:36}) are no 
longer valid since the concept of a self-regulating cluster goes down the drain 
in such peculiar situation. In the currently proposed code, when the search 
turns out to be non-convergent, the query distance from the center of mass of 
the outgoing kNN cluster at each iteration is recorded in a history array. The 
solution with the shortest distance to the center of mass is chosen as soon as 
the history matrix becomes periodic. Obviously, when this occurs, the best 
neighborhood found is not a self-regulating cluster, since the region subtended 
by the ellipsoidal hull is populated asymmetrically. This will cause edge 
effects in the SPH simulation but a similar situation would already occur with 
the spherical (isotropic) neighborhood.

The process above described requires many iterations, about twice the number of 
iterations needed to find a self-regulating cluster. The same considerations 
are used if the query is marginally out of distribution. Fortunately, the vast 
majority of kNN clusters found by the just exposed algorithm are 
self-regulating. This is an important fact since the vast majority of particles 
will be the center of a self-regulating cluster. Self-regulating clusters very 
accurately represent local trends in multivariate mass distribution in the 
anisotropic SPH simulation. Thus, it is likely that a particle is at the center 
of mass of its anisotropic neighborhood, favoring a fair interpolation with an 
ellipsoidal kernel, as stated by the multivariate analysis 
\citep{DudaHartStork:2001}.

The reader should bear in mind that each search for an optimal kNN cluster, be 
it the self-regulating one or the one with the shortest distance from the query, 
requires iterative searches through the covariance octree, which is in general 
$O(kL\log{N})$ \citep[e.g.,][]{Marinho-Andreazza:2010}, where $N=|\mathcal{D}|$, 
and $L$ is the average number of iterations (usually not much greater than $10$) 
to find a self-regulating cluster. In practice, a tolerance can be used for the 
thickness of the ellipsoidal region, studied in the next section, if the third 
principal component is very small or zero. In such a situation it becomes 
impossible to compute the kernel gradient, c.f. Sec.~\ref{sec:04}. One possible 
criterion to limit the thickness is that $\sigma_3$ cannot be less than half 
the estimated inter-particle distance, $\lambda=\sigma_1{k}^{-1/3}$, where 
$\sigma_1$ is the major magnitude of the principal components and $k$ is the 
preset number of nearest neighbors. Alternatively, the number of iterations can 
be a counter whose maximum $L_{\max}$ limits the volume thickness, $\sigma_3$. 
For example, adopting $L_{\max}=4$ has shown excellent results for the minimum 
thickness of the ellipsoidal kNN region. In this case, it is possible that we 
have an approximate self-regulating kNN cluster in place of the exact ones. 
Still, the experience has shown that several exact self-regulating clusters 
occur even within the $L_{\max}=4$ tolerance limit.

Since not all kNN clusters shall be exactly self-regulating, we will refer to 
them hereafter as simply a kNN cluster with the presumption that they are 
mostly good approximations to the ideal self-regulating kNN clusters. 

After a cycle of time integration, ulterior search for new self-regulating kNN 
is very efficient once too few of the neighborhood has changed due to the small 
magnitude of the time step (c.f. Sec.~\ref{sec:08}).

\section{The ellipsoidal hull for the self-regulating kNN cluster and the 
smoothing tensor}\label{sec:03}

The convex hull of the self-regulating kNN cluster is the smallest ellipsoid 
proportional to the confidence ellipsoid. In other words, it is a scale change 
in the confidence ellipsoid, maintaining the proper aspect ratio, and having 
the most distant neighbor in the kNN cluster on its surface. The query $q$ 
itself is the geometric center of the ellipsoidal hull. Alternatively, such an 
envelope corresponds to the smallest sphere with radius $\zeta_\text{max}(q)$ in 
the uncorrelated space $\vecspc{U}$, where $\zeta_\text{max}(q)$ is computed as 
the maximum Mahalanobis distance amongst the $k$-nearest neighbors:
\begin{equation}
\zeta_\text{max}(q)=
\sqrt{
\max_{p\in\mathcal{C}_k(q)}
    \bigset{
        \zeta_p^2=
        (\vec{r}_p-\vec{r}_{q})^t
        \tensor\Sigma_{\mathcal{C}_k(q)}^{-1}
        (\vec{r}_p-\vec{r}_{q})
    }}
\end{equation}
where $\vec{r}_{q}$ is the query position and $p\ne{q}$ stands for some of the 
proper $k$-nearest neighbors of $q$. Thus, the kNN-cluster's convex hull is the 
ellipsoid whose equation is written as
\begin{equation}
        (\vec{r}_p-\vec{r}_{q})^t
        \frac1{\zeta_\text{max}^2}
        \tensor\Sigma_{\mathcal{C}_k(q)}^{-1}
        (\vec{r}_p-\vec{r}_{q})
        =1
\end{equation}
which can be rewritten as
\begin{equation}\label{eq:38}
        (\vec{r}_p-\vec{r}_{q})^t
        \tensor{H}^{-2}
        (\vec{r}_p-\vec{r}_{q})
        =1
\end{equation}
where it is defined the {\em smoothing tensor}, $\tensor{H}$, given
the kNN cluster, $\mathcal{C}_k(q)$ and the query particle $q$:
\begin{equation}\label{eq:39}
        \tensor{H}
        \equiv
        \zeta_\text{max}\tensor\Sigma_{\mathcal{C}_k(q)}^{1/2}
\end{equation}
whose spectral decomposition, known the
$\tensor\Sigma_{\mathcal{C}_k(q)}$ eigenvalues/eigenvectors, is given by
\begin{equation}
        \tensor{H}
        =
            h_1\vec{e}_1\vec{e}_1^t +
            h_2\vec{e}_2\vec{e}_2^t +
            h_3\vec{e}_3\vec{e}_3^t
\end{equation}

The $\tensor{H}$'s eigenvalues, the {\em principal smoothing lengths}, 
are given in terms of the principal components according to the following scale 
change:
\begin{equation}
h_1=\zeta_\text{max}\sigma_1,\;
h_2=\zeta_\text{max}\sigma_2,\;
h_3=\zeta_\text{max}\sigma_3
\end{equation}

The quadratic form expressed on the left-hand side of equation~(\ref{eq:38})
induces the definition of the $\tensor{H}$-normalized, particle to query 
distance according to the
following equation:
\begin{equation}\label{eq:42}
        \delta_p=\sqrt{
        (\vec{r}_p-\vec{r}_{q})^t
        \tensor{H}^{-2}
        (\vec{r}_p-\vec{r}_{q})}
\end{equation}
where $p$ is the generic particle index and $\tensor{H}^{-1}$, the inverse of 
the smoothing tensor, can be easily computed as
\begin{equation}\label{eq:44}
        \tensor{H}^{-2}
        =
            \frac{\vec{e}_1\vec{e}_1^t}{h_1} +
            \frac{\vec{e}_2\vec{e}_2^t}{h_2} +
            \frac{\vec{e}_3\vec{e}_3^t}{h_3}
\end{equation}

Of course, $\tensor{H}$ has the same normalized eigenvectors as does 
$\tensor\Sigma_{\mathcal{C}_k}(q)$. For brevity, we call hereafter the 
$\tensor{H}$-normalized distance as simply $\tensor{H}$-distance. Thus, the 
outermost neighbor in ${\mathcal{C}_k}(q)$ is the particle whose query's 
$\tensor{H}$-distance is exactly $\delta_p=1$.

The tensor $\tensor{H}$ spans another uncorrelated vector space, called 
hereafter the smoothing space $\vecspc{S}_{q}$, whose origin corresponds to the 
query position $\vec{r}_q$, namely,
\begin{equation}
 \vecspc{S}_{q}=
 \bigset{
    \vec\xi=\tensor{H}^{-1}
        (\vec{r}-\vec{r}_{q}),
    \;\forall\vec{r}\in\vecspc{E}
 }
\end{equation}

Now, we call the kNN cluster ${\mathcal{C}_k}(q)$ as simply the smoothing 
cluster. Analogously, the region comprising the convex hull, 
$\tensor{H}$-ellipsoid, is called the smoothing region, which is in general the 
region corresponding to the inner region of the support of a compact-support 
smoothing kernel, $K:\vecspc{S}_{q}\mapsto\vecspc[]{R}$. Of course, the compact 
support corresponds to the unit sphere $\mathcal{S}_1$ in the smoothing space 
$\vecspc{S}_{q}$ so that $K(\vec\xi)=0$ if and only if $|\vec\xi|\ge1$. On the 
other hand, $K(\vec\xi)\ne0$ if and only if $|\vec\xi|<1$, which requires the 
kernel to be positive inside the smoothing region $\mathcal{S}_1$. The query 
particle is generally called a smoothed particle, and its neighboring 
particles, within the smoothing cluster, are called smoothing particles.
When writing the SPH equations of motion, even in the traditional isotropic 
approach, we should extend the idea of a smoothing cluster no longer to the kNN 
cluster but its symmetric closure. In this case, the smoothing cluster is a 
superposition of the original kNN cluster plus the set of particles that 
consider the query itself as one of its $k$-nearest neighbors. Such an 
extension is necessary to have the equations of motion conserving both momentum 
and total energy \citep[e.g.,][]{Monaghan:1992,HernquistKatz:1989}.

\section{Smoothing kernel}\label{sec:04}

In this section, the anisotropic smoothing kernel model will be discussed. We 
will show that the concept of smoothing space reduces the interpolation problem 
to the traditional case of isotropic interpolation. This is because the kNN 
cluster has a spherical outline in the smoothing space $\vecspc{S}_{q}$. Thus, 
everything that is done in this space corresponds to making the interpolations 
using an ellipsoidal kernel in the original simulation space $\vecspc{E}$.

The anisotropic smoothing kernel, 
$W_\tensor{H}:\vecspc{E}\mapsto\vecspc[]{R}_+$, can be conveniently
defined in terms of a dimensionless smoothing function, 
$K:\vecspc{S}_q\mapsto\vecspc[]{R}_+$, as already mentioned at the end of the 
previous section. 

The anisotropic smoothing kernel $W_\tensor{H}$ can be conveniently
defined in terms of a spherical and dimensionless smoothing function $K$, 
regarding equation~(\ref{eq:44})
\begin{equation}
W_\tensor{H}(\vec{r})
=
\frac{1}{\det\tensor{H}}K(\tensor{H}\vec{r}),
\end{equation}
as already mentioned at the end of the previous section. The smoothing kernel 
is a non-negative function whose domain is the original simulation space, 
namely, $W_\tensor{H}:\vecspc{E}\mapsto\vecspc[]{R}_+$. While the kernel 
function is spherical and defined in the smoothing space, 
$K:\vecspc{S}_q\mapsto\vecspc[]{R}_+$. As a rule, we adopt the kernel function 
$K$ as having compact support, defined as the unit sphere $\mathcal{S}_1$ 
centered on the origin, in the smoothing space $\vecspc{S}_q$. Consequently, 
the smoothing kernel is also a compact support function, whose support is the 
ellipsoid centered on the query particle. To recall, the ellipsoid semi-major 
axes are defined by the eigenvectors times the respective eigenvalues of the 
smoothing tensor.

The kernel model adopted in the present work was the 3D B-spline kernel 
\citep[e.g.,][]{MonaghanGingold:1983} shown in \ref{app:A}.

Since the kernel function $K$ is presumed spherically symmetric in the 
smoothing space $\vecspc{S}_{q}$, we can write the kernel effect of particle 
$q$ over particle $p$ as
\begin{equation}\label{eq:49}
W_\tensor{H}(\vec{r}_p-\vec{r}_{q})
=
{\det\tensor{H}^{-1}}K(\xi_{pq})
=\frac1{h_1h_2h_3}K(\xi_{pq})
\end{equation}
where,
\begin{equation}\label{eq:50}
 \xi_{pq}=
        \bigl|\tensor{H}^{-1}(\vec{r}_p-\vec{r}_{q})\bigr|
        =
        \sqrt{
        \xi_1^2 + \xi_2^2 + \xi_3^2}
\end{equation}
with
\begin{equation}\label{eq:50a}
    \xi_j=\frac{\vec{e}_j\cdot(\vec{r}_p-\vec{r}_{q})}{h_j},
    \;j=1,2,3
\end{equation}

The kernel normalization condition requires that
\begin{equation}\label{eq:51}
    \int_{\vecspc{E}}
        W_\tensor{H}(\vec{r})dx^3=1
\end{equation}
which, in spherical coordinates, regarding the spherical symmetry of the
kernel function $K$, it is equivalent to
writing
\begin{equation}
 \int_0^1
    \xi^2K(\xi)d\xi=\frac1{4\pi}
\end{equation}

Another useful formulation for the kernel function is the Cartesian-separable 
kernel
function in terms projections product of the individual function of projections 
in the principal directions,
$\vec{e}_1$, $\vec{e}_2$, $\vec{e}_3$, namely
\begin{equation}\label{eq:53}
 W_\tensor{H}(\vec{r})=
    \frac{K(\xi_1)}{h_1}
    \frac{K(\xi_2)}{h_2}
    \frac{K(\xi_3)}{h_3}
\end{equation}
where $\xi_j=\vec{r}\cdot\vec{e}_j/h_j$, $j=1,2,3$.
Thus, one finds, regarding equation~(\ref{eq:51}), that
\[
    \int_{\vecspc{E}}
        W_\tensor{H}(\vec{r})dx^3=
    \int_{-1}^1K(\xi_1)d\xi_1
    \int_{-1}^1K(\xi_2)d\xi_2
    \int_{-1}^1K(\xi_3)d\xi_3=1\]
which means
\begin{equation}
    \int_0^1K(\xi)d\xi=\frac12
\end{equation}
where it is presumed that $K(\xi)=K(-\xi)$ to ensure the symmetrical behavior
of the kernel interpolation technique discussed later.

In order to perform the SPH interpolation equations, the knowledge of the kernel
gradient $\vec\nabla{W}_\tensor{H}$ is required. If one adopts the kernel
formulation~(\ref{eq:49}), one has from (\ref{eq:50}) and (\ref{eq:50a}) the
following equation
\begin{equation}\label{eq:54}
 \vec\nabla{W}_\tensor{H}(\vec{r})=
 \det\tensor{H}^{-1}
    \frac1\xi{K^\prime(\xi)}\tensor{H}^{-2}\vec{r}
\end{equation}
where $$\xi=|\tensor{H}^{-1}\vec{r}|$$
and
$$K^\prime(\xi)=\deriv[\xi]{K(\xi)}$$

If the kernel is written in terms of separable
functions as in equation~(\ref{eq:53}), one finds
\begin{equation}
 \vec\nabla{W}_\tensor{H}(\vec{r})=
 W_\tensor{H}(\vec{r})
 \biggl[
 \frac{K^\prime(\xi_1)}{K(\xi_1)}\frac{\vec{e}_1}{h_1}
 +
 \frac{K^\prime(\xi_2)}{K(\xi_2)}\frac{\vec{e}_2}{h_2}
 +
 \frac{K^\prime(\xi_3)}{K(\xi_3)}\frac{\vec{e}_3}{h_3}
 \biggr]
\end{equation}
which requires the following regularity condition
$$\lim_{\xi\to1}K^\prime(\xi)/K(\xi)=
K^\prime(1)/K(1)=0.$$

It is adopted in the present work the formulations given in
equations~(\ref{eq:49}) and (\ref{eq:54}).

Still, with respect to equation~(\ref{eq:49}), it will be useful later to know
the partial derivatives of the smoothing kernel with respect to the principal
smoothing lengths. Thus,
\begin{equation}\label{eq:58}
\pderiv[h_j]{W_\tensor{H}(\vec{r})}=
\frac{-1}{h_1h_2h_3}
\biggl[
\frac{K(\xi)+K^\prime(\xi)\xi}{h_j}
\biggr];\;\;j=1,2,3
\end{equation}

If the smoothing tensor is computed from the point of view of the $q$ particle,
namely $\tensor{H}=\tensor{H}_q$, then $q$ is the center of the smoothing
region. If the $p$ particle is outside that region, it gives no contribution to
interpolation procedure. Similar reasoning occurs with respect to the particle
$p$ being the center of the interpolation. However, it is easy to show that
assuming a single point of view for only one of the particles violates the
conservation of momentum, whether linear or angular. such a symmetry issue
requires the adoption of a symmetrizing kernel. It is adopted in the present
work the \citet{HernquistKatz:1989} gather-scatter approach, namely
\begin{equation}\label{eq:59}
 \overline{W}_{pq}\equiv 
    .5\bigl[
        W_{\tensor{H}_q}(\vec{r}_p-\vec{r}_{q})
        +
        W_{\tensor{H}_p}(\vec{r}_p-\vec{r}_{q})
    \bigr]
\end{equation}

In order to quickly access the list of particles that make an effective
contribution to the smoothing kernel in symmetrized form above, it is necessary
to define the list of {\em effective neighboring} particles, which is defined 
as the {\em symmetric closure} of the kNN relation shown below.

Let $\mathcal{N}_k:\mathcal{D}\times\mathcal{D}$ be the kNN relation defined in 
the dataset $\mathcal{D}$ as earlier defined, so that, for all query particle 
$q\in\mathcal{D}$, we have the $(k+1)$-element set $\mathcal{N}_k(q)=\set{q, 
p_1, \ldots, p_k}$ ordered by distances to the query $q$: 
$\delta(q,p_{j})< \delta(q,p_{j+1})$, for all $j$ in $1<j<k$. As the reader is 
aware, the kNN relation is not symmetric. However, its symmetric closure
$\mathcal{N}_k^{*}$ can be defined as follows \citep[see, e.g.,][]{Marinho:2014}
\begin{equation}\label{eq:60}
 \mathcal{N}_k^{*}(p)\equiv \mathcal{N}_k(p)
 \cup \set{q\in\mathcal{D}\;|\;p\in\mathcal{N}_k(q)}
\end{equation}
The above definition for symmetric closure is consistent with the fact that
\begin{enumerate}
 \item $q\in\mathcal{N}_k^{*}(q)$ (reflexive);
 \item $r\in\mathcal{N}_k^{*}(q) \;\wedge\; 
q\in\mathcal{N}_k^{*}(p) \Rightarrow r\in\mathcal{N}_k^{*}(p)$ (transitive);
 \item $q\in\mathcal{N}_k^{*}(p) \Leftrightarrow p\in\mathcal{N}_k^{*}(q)$ 
(symmetric).
\end{enumerate}

\begin{algorithm}
\caption{The symmetric closure relation}
\begin{algorithmic}[1]
 \Require a $\mathcal{N}_k\subseteq\mathcal{D}\times\mathcal{D}$ relation
 \Ensure a symmetric closure
$\mathcal{N}_k^{*}\subseteq\mathcal{D}\times\mathcal{D}$ of
$\mathcal{N}_k$
\State $\mathcal{N}_k^{*}\gets\emptyset$
\ForAll{$p\in\mathcal{D}$}
    \ForAll{$q\in\mathcal{N}_k(p)$}
        \State \textbf{add} $q$ \textbf{to} $\mathcal{N}_k^{*}(p)$
        \If{$p\notin\mathcal{N}_k(q)$}
            \State \textbf{add} $p$ \textbf{to} $\mathcal{N}_k^{*}(q)$
        \EndIf
    \EndFor
\EndFor
\end{algorithmic}\label{alg:01}
\end{algorithm}

For a better understanding of the definition~(\ref{eq:60}), it is introduced 
the Algorithm \ref{alg:01} that builds a symmetric closure, given the kNN 
relation, $\mathcal{N}_k$, as input. A relation involving two categories is 
represented as a table. This is a reason for the initialization 
$\mathcal{N}_k^{*}\gets\emptyset$ in line 1 of the algorithm since lines 4 and 
6 adds iteratively members to the symmetric closure. The algorithm has time 
complexity $O(Lk^2N)$, with $L$ being the mean number of required iterations 
and presuming that the number of nearest neighbors $k$ is input by the user, 
and that it is in principle independent on the total number of particles $N$ in 
the dataset $\mathcal{D}$. If $k\propto\sqrt{N}$, then we have the time 
complexity given by $O(LN^2)$. 

The kNN algorithm to assemble the relation $\mathcal{N}_k$ is omitted here but 
it has time complexity $O(kN^2\log{N})=O(kN^3)$, which gives $O(N^{3.5})$ if $k$ 
is chosen to be $k\propto\sqrt{N}$.

\section{Covariance-based anisotropic SPH}\label{sec:05}

This section is dedicated to applying the self-regulating kNN cluster estimate, 
studied in the previous sections, to smoothed particle hydrodynamics (SPH). 
What matters here is the shape of the smoothing kernel's compact support, which 
is defined according to the ellipsoidal hull as previously discussed in 
Sections~\ref{sec:03} and \ref{sec:04}.

Hereafter, it is assumed that the reader is familiar with the isotropic SPH 
formalism to understand how we convert the SPH equations of motion to the 
anisotropic interpolation methodology. However, if there is a need to review 
the subject, it is very recommended \cite{Monaghan:1992} and 
\cite{Monaghan:2012}.

The modeled gas is assumed to be a compressible, non-viscous,
self-gravitating fluid. It is also assumed that the fluid is both chemically and
nuclear inert and that there are no sources or sinks of matter nor heat. In 
these terms, the Lagrangian fluid conservation equations are written as follows.

The continuity equation, aka mass conservation equation, is written as
\begin{equation}\label{eq:62}
 \dot{\rho}+\rho\vec\nabla\cdot\vec{v}=0
\end{equation}

Linear momentum conservation equation,
\begin{equation}
 \dot{\vec{v}}=
    -\frac{\vec\nabla{P}}\rho
    -\vec\nabla\phi
\end{equation}
where $-\vec\nabla\phi$ is the gravity acceleration caused by the entire fluid
over a co-moving differential fluid element located at position $\vec{r}$, with
velocity $\vec{v}$.

The thermodynamics first law, or simply the energy conservation equation, for
the adiabatic regime, is written as
\begin{equation}
 \dot{u}=-\frac{P}\rho\vec\nabla\cdot\vec{v}
\end{equation}
where the right-hand side expresses the adiabatic compression heat, if
$\vec\nabla\cdot\vec{v}<0$, or the adiabatic expansion cooling, if
$\vec\nabla\cdot\vec{v}>0$.

Essentially, the equations of motion above must be translated into the SPH
discrete interpolation formulas. These particle interpolations have as a
trade-off the reduction in spatial resolution for replacing the
continuum of the distribution theory with the discrete and using varying
smoothing lengths; in the present case, the principal components of
the smoothing tensor. Still, many, if not all, SPH works essentially use these
rather simple interpolation formulas. Let us start with the SPH density, which
is essentially a density estimation technique 
\citep[e.g.,][]{DudaHartStork:2001}.

The anisotropic density estimate on the particles in the data set
$\mathcal{D}$ is computed as usual, just as it would be computed for the
isotropic case, since that the anisotropy is embedded in the smoothing kernel.
To know, the summation interpolant for anisotropic density has the same aspect
as in the isotropic approach,
\begin{equation}\label{eq:61}
 \rho_p = \sum_{q\in\mathcal{N}_k^{*}(p)} m_q\overline{W}_{pq}
\end{equation}
with $\overline{W}_{pq}$ written as in equation~(\ref{eq:59}) and
$\mathcal{N}_k^{*}(p)$ is the effective neighboring cluster to the query
particle $p$, as defined as the symmetric closure of the kNN in
equation~(\ref{eq:60}) and also in Algorithm~\ref{alg:01}.

From the SPH density equation~(\ref{eq:61}), one finds that the particle's
effective volume is estimated as
\begin{equation}
 V_p=\frac{m_p}{\rho_p}
\end{equation}
Consistently, the total mas of the dataset $\mathcal{D}$ is given by
\begin{equation}
 M=\sum_{p=1}^N
 \rho_pV_p
 =
 \sum_{p=1}^Nm_p
\end{equation}

On the other hand, the kernel normalization condition requires that
\begin{equation}
 \sum_{q\in\mathcal{N}_k^{*}(p)}
 V_q\overline{W}_{pq}=
 \sum_{q\in\mathcal{N}_k^{*}(p)}
 \frac{m_q}{\rho_q}\overline{W}_{pq}
 =
 1
\end{equation}
Let $\nu_p$ be the normalization check estimated on the $p$ particle:
\begin{equation}
 \nu_p=\sum_{q\in\mathcal{N}_k^{*}(p)}
 \frac{m_q}{\rho_q}\overline{W}_{pq}
\end{equation}
Thus, the accuracy of the interpolation equations indirectly depends on having
$\nu_p=1$ for all $p\in\mathcal{D}$. It can be concluded from the 
analysis made by \cite{HernquistKatz:1989} that $\nu_p=1+O(h_p^2)$ for the 
isotropic case. So, one can infer that in the present case the normalization 
error goes with the square of the major principal component of the smoothing 
tensor.

In order to check the consistency of equation~(\ref{eq:62}) with SPH 
interpolation scheme, it is necessary to estimate the total derivative of the 
interpolated density given in equation~(\ref{eq:61}),
\begin{equation}\label{eq:62a}
 \dot\rho_p = \sum_{q\in\mathcal{N}_k^{*}(p)}
 m_q
 \deriv{\overline{W}_{pq}}
\end{equation}
so that, expanding the right-hand side of the above equation, we have
\begin{equation}\label{eq:63}
 \dot\rho_p
 =
 \sum_{q\in\mathcal{N}_k^{*}(p)}
 m_q
 \vec\nabla_p\overline{W}_{pq}\cdot\vec{v}_{pq}
 +
 \sum_{q\in\mathcal{N}_k^{*}(p)}
 m_q D_{pq}
\end{equation}
where $\vec{v}_{pq}=\vec{v}_p-\vec{v}_q$ is the approach velocity of
the $p$ particle toward the $q$ particle, and the kernel gradient is computed 
as in equation~(\ref{eq:54}) adapted to the gather-scatter form given in 
equation~(\ref{eq:59}).

The coefficient $D_{pq}$ is defined here as the kernel-diffusion coefficient, 
involving the pair $(p,q)$, which is written as
\begin{equation}
  D_{pq}=
 \sum_{j=1}^3
 \biggl[
    \dot{h}_{j,p}\frac{\partial{\overline{W}_{pq}}}{\partial{h}_{j,p}}
    +
    \dot{h}_{j,q}\frac{\partial{\overline{W}_{pq}}}{\partial{h}_{j,q}}
 \biggr]
\end{equation}
with $h_{j,p}$ being the $j$th principal smoothing length, having $p$ as query,
and symmetrically $h_{j,q}$ is the $j$th principal smoothing length having 
the $q$-particle as query. The $h_j$-partial derivatives of the smoothing 
kernel is computed according to equation~(\ref{eq:58}).

The kernel-diffusion effect is a consequence of having the $\tensor{H}$ 
components varying along with the fluid flow 
\citep{OwenVillumsenShapiroMartel:1998}, which is generally ignored in the vast 
majority of the SPH literature, even using isotropic smoothing length 
\citep[e.g.,][]{MarinhoLepine:2000}. However, the effects of varying smoothing 
lengths in the isotropic formulation were discussed by \cite{Monaghan:1992}. As 
previously remarked in the previous paragraph, \cite{HernquistKatz:1989} has 
shown that the interpolation errors are of $O(h^2)$, which are greater than the 
errors of a variety of time-integration schemes adopted in SPH. For instance, 
the second-order accuracy leapfrog has truncation error of roughly 
$O(\delta{t}^3)=O(h^3/v^3)$, where $\delta{t}$ is a time step scale and $v$ is 
a dynamical scale of velocities for an SPH simulation.

The first summation, from left to right in the right-hand side of 
equation~(\ref{eq:63}), is the negative of the smoothed velocity divergence of 
the $p$ particle, multiplied by its density [see equation~(\ref{eq:62})]. 
To know, the summation interpolant to have the smoothed velocity divergence is 
written as
\begin{equation}\label{eq:64}
 \vec\nabla_p\cdot\vec{v}_p
 =
 -\frac1{\rho_p}\sum_{q\in\mathcal{N}_k^{*}(p)}
 m_q
 \vec\nabla_p\overline{W}_{pq}\cdot\vec{v}_{pq}
\end{equation}

Suggestively, we rewrite equation~(\ref{eq:63}), using equation~(\ref{eq:64}),
in order to remember the Lagrangian continuity equation with a source term in
the right-hand side. Namely,
\begin{equation}\label{eq:65}
 \dot\rho_p + \rho_p\vec\nabla_p\cdot\vec{v}_p
 = \sum_{q\in\mathcal{N}_k^{*}(p)} m_q D_{pq}
\end{equation}
to express the mass conservation error. Thus, we can interpret the kernel
diffusion term as an error in estimating the total density derivative by just
considering the actual mass continuity equation~(\ref{eq:62})
rather than equation~(\ref{eq:65}).

Although we do not estimate the terms of diffusion of the kernel in the present 
work, we are aware that such an error does exist. In prolonged simulations, it 
can affect the results of the SPH equations that explicitly depend on the 
divergence of speed, as is the case of the energy conservation equation 
discussed later.

With the neglect of the kernel diffusion term, we have the following 
approximation
\begin{equation}\label{eq:65a}
 \dot\rho_p = 
\sum_{q\in\mathcal{N}_k^{*}(p)}m_q\vec\nabla_p\overline{W}_{pq}\cdot\vec{v}_{pq}
\end{equation}

The relevant SPH equations to perform the anisotropic SPH simulation of an ideal
adiabatic gas are written as
\begin{equation}\label{eq:77}
 \dot{\vec{v}}_p=
 -\sum_{q\in\mathcal{N}_k^{*}(p)}
 m_q
 \vec\nabla_p\overline{W}_{pq}
 \biggl(
 \frac{P_p}{\rho_p^2}+
 \frac{P_q}{\rho_q^2}
 +\Pi_{pq}
 \biggr)
 -\vec\nabla_p\phi_p
\end{equation}
where $\Pi_{pq}$ is the anisotropic artificial viscosity involving both shear
and bulk effects of the interacting particles $p$ and $q$, which will be
discussed later; $-\vec\nabla_p\phi_p$ is the gravity acceleration computed
over the $p$ particle. It is easy to show that such a formulation is
momentum-preserving since that
$\vec\nabla_p\overline{W}_{pq}=-\vec\nabla_q\overline{W}_{pq}$.

The adiabatic SPH equation for thermodynamics first law, regarding the kernel
symmetry above commented to allow energy conservation, is commonly written as
\begin{equation}\label{eq:78}
 \dot{u}_p=
 \frac12
 \sum_{q\in\mathcal{N}_k^{*}(p)}m_q\vec\nabla_p\overline{W}_{pq}\cdot
    \vec{v}_{pq}
    \biggl(
        \frac{P_p}{\rho_p^2}+
        \frac{P_q}{\rho_q^2}
        +\Pi_{pq}
    \biggr)
\end{equation}

The derivation of both equations~(\ref{eq:77}) and (\ref{eq:78}) is shown,
e.g., in \cite{Monaghan:1992}.

At the end of this section, it is clear that the only thing that changes in the 
adaptive anisotropic SPH is the mathematical modeling of the smoothing kernel 
and its spatial derivatives. The final aspect of the basic SPH equations of 
motion remains unchanged. This makes it easier to perform tests on possible 
anisotropic kernel models to be "plugged in" without having to make any changes 
to the formal SPH equations.

\section{Anisotropic artificial viscosity}\label{sec:06}

This is perhaps the most critical part of designing an anisotropic SPH code, 
which is the anisotropic model of artificial viscosity. There are several 
models 
in the literature, but here a variant of the viscosity by 
\cite{Marinho-et-al:2001}, which, at that time, was an adaptation of Monaghan's 
artificial viscosity  \citep{Monaghan:1992} to the magnetic stress tensor. The 
isotropic Monaghan's artificial viscosity is sensitive to the adiabatic speed 
of 
sound, which requires a brief review of the thermodynamics of ideal gas. For 
this reason, this matter deserves to be highlighted in a separate section so 
that details of the transcription from Monaghan's isotropic model to the 
present 
anisotropic one are studied with special attention.

\subsection{Thermodynamics considerations}

The thermodynamic state of an ideal gas can be written as
\begin{equation}\label{eq:71}
 \frac{P}\rho=\frac{RT}{\bar\mu}
\end{equation}
where $R$ is the gas constant ($0.08206$ L atm K$^{-1}$mol$^{-1}$),
$\bar\mu$ is the mean molecular weight of one mole of the gas mixture, say the
mass in grams of $6.02214076\times10^{23}$ molecules. For example, $\bar\mu=2$
g mol$^{-1}$ for molecular hydrogen, H$_2$.

The specific thermal energy (erg g$^{-1}$) is given by 
\begin{equation}\label{eq:72}
 u=\frac\phi2\frac{RT}{\bar\mu}=\frac\phi2\frac{P}\rho
\end{equation}
where $\phi$ is the average number of degrees of freedom of the 
molecular/atomic mixture, which is known by statistical mechanics to be 
approximately $\phi=3$ for a monatomic gas, $\phi=5$ for diatomic gas, and 
$\phi=6$ for non-linear molecules with the neglect of the internal modes of 
vibration.

The temperature in Kelvin can be useful to monitor what is going on with the 
gas temperature during the simulation steps, and it is computed from 
equations~(\ref{eq:71}) and (\ref{eq:72}), yielding
\begin{equation}\label{eq:82}
 T=\frac{2}{\phi}\frac{\bar\mu}{R}u
\end{equation}

The adiabatic index $\gamma$ can be written in terms of the normalized heat 
capacities at constant pressure $C_p$ and at constant volume $C_v$ as well 
as in terms of the mean degrees of freedom
$\phi$ as
\begin{equation}\label{eq:78a}
 \gamma=\frac{C_p}{C_v}=1+\frac1{C_v}=1+\frac2\phi
\end{equation}
so that, from equation~(\ref{eq:72}), one has the equivalent form of 
equation~(\ref{eq:71}), namely,
\begin{equation}\label{eq:83}
 \frac{P}\rho=(\gamma-1)u
\end{equation}
which is useful to calculate pressures in equations~(\ref{eq:77}) and 
(\ref{eq:78}), for the ulterior integration scheme as will be shown in 
Section~\ref{sec:08}.
The quantities $C_p$ and $C_v$ are related according to
\begin{equation}
 C_p=C_v+1
\end{equation}
From equation~(\ref{eq:78a}), one finds
\begin{equation}
 C_v=\frac\phi2
\end{equation}

One can easily derive, by means of the differential adiabatic transformation, 
the well-known equation
\begin{equation}\label{eq:84a}
 {P}={P_0}\biggl(\frac{\rho}{\rho_0}\biggr)^\gamma
\end{equation}
For a monatomic gas we have $\gamma=5/3=1.67$, and for a diatomic gas
we have $\gamma=1.4$.

In the case of a polytrope, one can rewrite the latter equation as
\begin{equation}\label{eq:84b}
 {P}=K\rho^{1+\frac1n}
\end{equation}
where $K$ is an arbitrary constant of proportionality and $n$ is known as the 
polytropic index. Such equation of state is known in astrophysics as the 
solution of the Lane--Endem equation \citep[e.g.,][]{Binney-Tremaine:1987}, 
who studied self-gravitating polytropic gas spheres. To have an idea of how 
equation~(\ref{eq:84b}) differs from (\ref{eq:84a}) we compare the adiabatic 
index with the polytropic exponent, yielding
\begin{equation}\label{eq:86a}
\gamma=1+\frac1{n}\;\;\;\;\text{or}\;\;\;\;n=\frac1{\gamma-1}
\end{equation}
from which we have the polytropic index $n=2.5$ for $\gamma=1.4$ (diatomic 
gas). A polytrope with index $n = 3$ ($\gamma=1.333$) is used to model 
main-sequence stars, corresponding to the Eddington standard model of stellar 
structure \citep[e.g.,][]{Mestel:2004}. A polytrope with 
index $n = 1.5$ is used to model fully convective star cores (as those of red 
giants), and also to model brown dwarfs and giant gaseous planets (like Jupiter 
and Saturn) \citep[e.g.,][]{Hansen-et-al:2004}. Neutron stars can be modeled as 
a polytrope with polytropic index $n=1$ 
\citep[e.g.,][]{Bera-et-al:2020,Kippenhan-et-al:2012}. The radial density 
profile of a polytrope changes from a bell-like curve (for higher polytropic 
indices) to a top-hat curve (for very small polytropic indices). Thus, the 
fluid becomes almost solid for very small values of the polytropic index, say 
$n\ll1$ (or $\gamma\gg1$). Anyway, the density profile $\rho=f(r)$ has a maximum 
in the center, $\rho_\text{max}=f(0)$, and decays to zero to a maximum radius 
$R$, say $f(R)=0$. The isothermal case occurs in the asymptotic behavior of the 
polytropic index $n\to\infty$. This is the case when pressure and density 
become proportional, $p=K\rho$.

One interesting approach is considering the case of a variant of polytrope with 
adjustable index $n$ according to
\begin{equation}\label{eq:82b}
 n(\rho) = 
    \frac{\phi_0}2
    \biggl(
      \frac{e^{\rho/\rho_c}}{e^{\rho/\rho_c}-1}
    \biggr)
\end{equation}
where $\phi_0$ is the average number of degrees of freedom of the gas mixture 
in the standard case as in equation~(\ref{eq:78a}), and $\rho_c$ is a critical 
density, which shall be chosen accordingly to the simulation scenario.
Thus, from equations~(\ref{eq:86a}) and (\ref{eq:82b}), one finds
\begin{equation}\label{eq:82c}
 \gamma = 1 + 
    \frac2{\phi_0}
    \bigl( 1 - e^{-\rho/\rho_c} \bigr)
\end{equation}
It is easy to show that the gas regime becomes weakly adiabatic when
$\rho\gg\rho_c$. Conversely, the regime is weakly isothermal when 
$\rho\ll\rho_c$.

Equation~(\ref{eq:82c}) is useful to simulate gas collapse to 
reproduce the formation of adiabatic core, when the interstellar medium 
changes from transparent to opaque, triggering an outward adiabatic shock, when 
changing from isothermal to adiabatic collapse.

The adiabatic sound speed for an ideal gas is given by
\begin{equation}\label{eq:84}
 c_s=\sqrt{\left(\frac{\partial{P}}{\partial\rho}\right)_s}
 =\sqrt{\gamma\frac{P}\rho}=\sqrt{\gamma(\gamma-1)u}.
\end{equation}
This is required by the mostly adopted model for isotropic artificial 
viscosity, which is the starting point for the anisotropic formulation discussed 
in the next subsection. Even in the multiphase model, c.f. 
equation~(\ref{eq:82c}), we have adopted the equation above to denote the 
quantity $c_s$ in the computation of the anisotropic artificial viscosity.

\subsection{The anisotropic artificial viscosity model}

The commonly adopted artificial viscosity shown in the SPH literature derives 
from the Monaghan formulation for the artificial viscosity term, $\Pi_{pq}$, 
appearing in the pressure-dependent equations of motion~(\ref{eq:77}) and 
(\ref{eq:78}) \citep[e.g.,][and references 
therein]{HernquistKatz:1989,Monaghan:1992,Monaghan:2012}, which is defined as
\begin{equation}\label{eq:76}
\Pi_{pq}=
\begin{cases}
\frac{-\alpha\mu_{pq}\bar{c}_{pq}+\beta\mu_{pq}^2}{\bar\rho_{pq}},
                 &\vec{v}_{pq}\cdot\vec{r}_{pq}<0\\
 0,              &\vec{v}_{pq}\cdot\vec{r}_{pq}\ge0
 \end{cases}
\end{equation}
where the over-barred quantities, say $\bar{a}_{pq}$ corresponds to the simple
arithmetic mean, $\bar{a}_{pq}=.5(a_p+a_q)$. On the other hand, the unbarred
vector quantities mean vector difference, say
$\vec{a}_{pq}=\vec{a}_p-\vec{a}_q$. The quantity $\mu_{pq}$ has the physical 
scale of velocity, and is defined as
\begin{equation}\label{eq:73}
        \mu_{pq}=
                \frac
                {\vec{v}_{pq}\cdot\vec{r}_{pq}/\bar{h}_{pq}}
                {|\vec{r}_{pq}|^2/\bar{h}_{pq}^2+\eta^2}
\end{equation}
where $\bar{h}_{pq}$ is the averaged smoothing length, used in the
conventional isotropic, density-adaptive SPH
\citep[see, e.g.,][]{HernquistKatz:1989}. The
coefficients $\alpha$ and $\beta$ are in general comparable to the unit 
\citep[e.g.,][]{Monaghan:1992}, and the shock-thickness term $\eta^2$ is 
set as $\eta^2=0.01$. Both $\alpha$ and $\beta$ must be fine tuned for 
shock simulations to avoid excessive shear effects and to not allow particle 
interpenetration.

The average adiabatic speed of sound, $\bar{c}_{pq}$, used in artificial
viscosity model in equation~(\ref{eq:76}) is computed from 
equation~(\ref{eq:84}) as follows
\begin{equation}
 \bar{c}_{pq}=
 \frac{\sqrt{\gamma(\gamma-1)u_p}+\sqrt{\gamma(\gamma-1)u_q}}2
\end{equation}

By inspecting equation~(\ref{eq:73}), it is suggestive that the anisotropic
artificial viscosity can be rewritten after doing the following replacements:
\begin{equation}
 \vec{r}_{pq}/\bar{h}_{pq}\longrightarrow
 \overline{\tensor{H}}^{-1}_{pq}\cdot\vec{r}_{pq}
\end{equation}
where
$\overline{\tensor{H}}_{pq}=.5(\tensor{H}_p+\tensor{H}_q)$,
and
\begin{equation}
 |\vec{r}_{pq}|^2/\bar{h}_{pq}^2\longrightarrow
 |\overline{\tensor{H}}^{-1}_{pq}\cdot\vec{r}_{pq}|^2\equiv
 \vec{r}_{pq}\cdot\overline{\tensor{H}}_{pq}^{-2}\cdot\vec{r}_{pq}
\end{equation}

The anisotropic artificial viscosity is introduced by modifying the
$\mu_{pq}$-velocity factor appearing in equation~(\ref{eq:76}), defined in 
equation~(\ref{eq:73}). Thus, equation~(\ref{eq:73}) is replaced by the 
following anisotropic velocity-scale model,
\begin{equation}\label{eq:80}
 \mu_{pq}^* =
 \frac
 {\vec{v}_{pq}\cdot\overline{\tensor{H}}_{pq}^{-1}\cdot\vec{r}_{pq}}
 {\vec{r}_{pq}\cdot\overline{\tensor{H}}_{pq}^{-2}\cdot\vec{r}_{pq}+\eta^2}
\end{equation}
From the latter result, one can see that equation~(\ref{eq:76}) is analogous to 
writing
\begin{equation}\label{eq:88}
\Pi_{pq}=
\begin{cases}
\frac{-\alpha\mu_{pq}^*\bar{c}_{pq}+\beta{\mu^*}_{pq}^2}{\bar\rho_{pq}},
&\vec{v}_{pq}\cdot\overline{\tensor{H}}_{pq}^{-1}\cdot\vec{r}_{pq}<0\\
 0,
&\vec{v}_{pq}\cdot\overline{\tensor{H}}_{pq}^{-1}\cdot\vec{r}_{pq}\ge0
 \end{cases}
\end{equation}

Examining equations~(\ref{eq:80}) and (\ref{eq:88}), one can see that the 
proposed anisotropic artificial viscosity depends not only on the relative 
approach velocity, $\vec{v}_{pq}\cdot\vec{r}_{pq}/|\vec{r}_{pq}|$, as in the 
Monaghan's model, but also depends on the directions relative to the principal 
components of the smoothing tensor, $\vec{p}_j=h_j\vec{e}_j,\,j=1,2,3$, 
whose anisotropic advance of the $p$-particle against to the $q$-particle is 
denoted by the double scalar product,
$$\vec{v}_{pq}\cdot\overline{\tensor{H}}_{pq}^{-1}\cdot\vec{r}_{pq}.$$
This is negative if the fluid is compressed anisotropically, i.e., compressed 
mainly against the smallest of the main directions, and positive for expansion, 
in a similar symmetry.
The bilinear form above reveals the artificial viscosity's anisotropic nature, 
denoted in equation~(\ref{eq:88}): the strongest shock component occurs 
preferentially against the plane whose normal vector is the smallest semi-major 
axis, namely $h_3\vec{e}_3$. Such reasoning stems from the fact that the 
particle distribution assumes an oblate ellipsoid molding the shock layer. 
Thus, 
the resulting artificial viscosity produces greater acceleration in the 
opposite direction to the shock and grows even more as the ellipsoid becomes 
more and more flattened. 

At first glance, the reader may find that both methods, the classic and the one 
proposed here, are equivalent. In fact, in the classic, there is a similar 
effect of the artificial viscosity reaction being intense when two particles 
approach. However, this effect is amplified in how the approach velocity is 
changed to the anisotropic form, mainly due to the denominator on the right 
side 
of the equation~(\ref{eq:80}).

\section{Gravity estimation}\label{sec:07}

Gravity acceleration was computed by a modified Tree-code method 
\citep{BarnesHut:1986}, where the covariance octree proposed by 
\cite{Marinho-Andreazza:2010} replaces the traditional octree. What changes is 
that, instead of the preset spatial tessellation of the computational space 
into 
cubes, or parallelepipeds, of the Barnes--Hut method, the Marinho--Andreazza's 
covariance octree allows a non-fixed geometry, based on the recursive division 
of the space by cutting planes, according to the principal components estimated 
over particles inside the dividing cells.

The cells resulting from covariance-based tessellation, given the distribution 
of SPH particles, are very similar to a 3D version of the images and plots 
shown 
in \cite{Marinho-Andreazza:2010}. Furthermore, the tree-descent algorithm to 
find the well-separated nodes is somehow similar to the algorithm to perform 
the 
anisotropic k-nearest neighbors, as shown in the just cited paper.

The article that describes in more detail and validates the gravity computation 
based on covariance octree is still in preparation. Even so, the full version 
code for performing anisotropic self-gravitating SPH simulations is available 
upon request.

To fulfill the tolerance condition \citep[e.g.][]{Appel:1985, BarnesHut:1986}, 
we established a similar criterion of a particle being well separated from a 
covariance-octree node according to the node's principal directions, as follows.

Firstly, consider the line-of-sight projections against the node's principal 
directions:
\begin{equation}
 \xi_j=
    \vec{e}_j\cdot
    \bigl(
        \vec{r}_{\nu}-\vec{r}_p
    \bigr);\;j=1,2,3
\end{equation}
where $\nu$ is the pointer (index) to the covariant-octree node at position 
$\vec{r}_{\nu}$, seen from the $p$-particle at position $\vec{r}_p$, which 
occurs along the tree-descent; $\vec{e}_j$ is the $j$th eigenvector for the 
covariance tensor $\tensor{\Sigma}_\nu$ evaluated from the $\nu$'s content.

Given the covariance eigenvalues of $\nu$ it is computed the principal areas of 
the minimal parallelepiped circumscribing the $\tensor{\Sigma}_\nu$-ellipsoid:
\begin{equation}
 S_1=\sigma_2\sigma_3,\;
 S_2=\sigma_1\sigma_3,\;
 S_3=\sigma_1\sigma_2
\end{equation}
whose normal vectors are the eigenvectors $\vec{e}_1$, $\vec{e}_2$ and
$\vec{e}_3$, respectively. For instance, the projected area $S_1^\prime$
against the line of sight of the particle $p$ is given by
\begin{equation}
S_1^\prime=
            \frac{|S_1\vec{e}_1\cdot(\vec{r}_{\nu}-\vec{r}_p)|}
            {|\vec{r}_{\nu}-\vec{r}_p|}
            =\frac{S_1|\xi_1|}{|\vec{r}_{\nu}-\vec{r}_p|}
\end{equation}
Then, the effective area seen from $p$ is computed as
\begin{equation}
 S_{\nu p}=
    {|\vec{r}_{\nu}-\vec{r}_p|}^{-1}
    \sqrt{\sum_{j=1}^3S_j^2\xi_j^2}
\end{equation}
so that the approximate solid angle seen from $p$, covering the node $\nu$, is 
computed as follows
\begin{equation}
 \Omega_{\nu p}=\frac{\pi\,S_{\nu p}}{|\vec{r}_{\nu}-\vec{r}_p|^2}
\end{equation}

The well-distant criterion is then expressed by the following predicate
\begin{equation}
 \Omega_{\nu p}\le\theta^2
\end{equation}
where $\theta^2$ is the square of the preset tolerance parameter. Thus, $p$ is 
well distant from $\nu$ if and only if the above predicate is true. It has been 
adopted $\theta=.25$ in the self-gravitating anisotropic SPH simulations shown 
in the test section.

\section{Time integration}\label{sec:08}

The SPH equations of motion were integrated using the adaptive leapfrog model
proposed by \citet{MarinhoLepine:2000}. After several tests with different
time-depth levels, we have adopted a time-depth equal to 12. This means that 
the leapfrog's binary scheduling hierarchy had the deepest time step of 1/4096 
the root time step. Larger time step particles are integrated first than the 
smaller ones. Paraphrasing, slower particles first.

Experience has shown that a small fraction of the entire data set reaches the 
deepest time steps for the test simulations presented here. Still, the number 
of 
time levels depends on the simulation purpose.

To set up the hierarchical multiple time step leapfrog scheme, it is necessary 
to assign a characteristic time to each particle according to the Courant 
stability criterion. Assuming the self-gravitating case of anisotropic SPH, one 
must estimate, for each particle, a gravitational time scale, a geometric time 
scale, for example, by dividing the shortest node length in the covariance tree 
divided by the particle speed, the thermodynamic time, involving artificial 
viscosity. The shortest of these times is assumed to be the characteristic time 
of the particle in question. For this time, an estimate is made by looking for 
the greatest power of 2 less than or equal to the characteristic time. This is 
the time that classifies the particle at the level of a binary integration 
tree, 
as proposed by \cite{MarinhoLepine:2000} and \cite{Marinho-et-al:2001}. The 
method in question vaguely resembles a hierarchical version of the round-robin 
scheduling model.

The adaptive time steps finite difference equations of motion are similar to 
those proposed by \cite{HernquistKatz:1989}. Such an integration scheme 
minimizes errors resulting from the particle having their individual time steps 
changing from one time-depth to another. As usual in the leapfrog scheme, 
positions must be delayed by half time step the initial conditions, while 
velocities have their initial conditions unchanged. Thus, after a number $n$ of 
integration cycles, velocities are at time level $n+1$ while positions are at 
the centered time level $n+1/2$. One difficulty occurs with quantities that 
depend not only on positions but also on velocities, as does the specific 
thermal energy rate and the artificial viscosity pressure. In this case, a 
temporary synchronism between velocities and positions must be made before 
computing these referred SPH quantities.

Particle positions are updated from time level $n-1/2$ to time level $n + 1/2$ 
according to the following second-order accurate equation:
\begin{equation}\label{eq:96}
\vec{x}_{n+1/2} = \vec{x}_{n-1/2} 
    + \vec{v}_{n}\overline{\tau}_{n}
    + \frac12\vec{a}_{n}\overline{\tau}_{n}\delta\tau_{n}
    + O[(\overline{\tau}_{n}+\delta\tau_{n})^3]
\end{equation}
where
\begin{equation}\label{eq:97}
 \overline{\tau}_{n}=.5\;(\tau_{n+1/2}+\tau_{n-1/2})
\end{equation}
is the midpoint time step between time levels $n-1/2$ and $n+1/2$, and
\begin{equation}\label{eq:98}
 \delta{\tau}_{n}=.5\;(\tau_{n+1/2}-\tau_{n-1/2})
\end{equation}
is the time step skew from time level $n-1/2$ to $n+1/2$, whereas 
$\tau_{n-1/2}$ and $\tau_{n+1/2}$ are the time steps at time levels $n-1/2$ and 
$n+1/2$, respectively.

Despite acceleration term is denoted as synchronized with velocities it can be 
written approximately depending velocities at time level $n$, but depending on 
positions at time level $n-1/2$, namely, 
$\vec{a}_{n}\cong\vec{a}(\vec{x}_{n-1/2}, \vec{v}_n)$. To recall from 
Sections~\ref{sec:05} and \ref{sec:07}, the acceleration 
vector is written as
\begin{equation}
 \vec a=-\frac{\vec\nabla{P}}{\rho}-\vec\nabla\Phi,
\end{equation}
which is the equivalent to the RHS of equation~(\ref{eq:77}).
It is important to remark that $P$ depends on the particle velocity due to the 
artificial viscosity model in equation~(\ref{eq:76}). 

To have velocity-synchronized positions it is necessary to perform the 
following prediction
\begin{equation}\label{eq:99}
\vec{x}_{n} = \frac{\vec{x}_{n+1/2} + \vec{x}_{n-1/2}}2
    + \frac18\vec{a}_{n}\tau_{n+1/2}^2 + O(\tau_{n+1/2}^3)
\end{equation}
It requires some iterations in correcting the equation~(\ref{eq:96}) more 
because of the pressure calculation than the gravity acceleration. However, I 
did not make this recurrence because this can be an excess of perfection and 
unnecessary overhead, doing a third-order accuracy correction. By the way, 
gravity computation is the fastest component of the present anisotropic, 
self-gravitating SPH code.

Velocities are position centered and are integrated as
\begin{equation}\label{eq:100}
\vec{v}_{n+1} = \vec{v}_{n} 
    + \vec{a}_{n+1/2}\;{\tau_{n+1/2}}+ O(\tau_{n+1/2}^3),
\end{equation}

Observing that 
\begin{equation}\label{eq:102}
\vec{v}_{n+1/2} = \frac{\vec{v}_{n+1} + \vec{v}_{n}}2
     + O(\tau_{n+1/2}^3),
\end{equation}
one can predict the position-synchronized velocities to get 
\[\vec{a}_{n+1/2}=\vec{a}(\vec{x}^{n+1/2},\vec{v}^{n+1/2}),\] which is 
necessary 
since the accelerations are the coefficients of the linear term in 
$\tau_{n+1/2}$ so that the equation~(\ref{eq:100}) can, in fact, have 
second-order accuracy, which requires some iterations. Such iterations can make 
the time of execution of the simulation relatively expensive. Each correction 
in 
the accelerations requires procedural calls for the calculation of the 
artificial viscosity, which requires a large number of tensor operations to the 
successive visits to the lists of effective neighbors. As previously commented, 
gravity is not so expensive due to covariance-octree descents' satisfactory 
performance given a well-chosen tolerance parameter.

Densities can be straightforwardly updated by exhaustively computing the 
anisotropic kNN. In this case, one has promptly that $\rho=\rho(\vec{x})$. To 
significantly reduce the computation time, one alternative approach is 
explicitly integrating equation~(\ref{eq:62}), namely,
\begin{equation}
 \dot{\rho}=-\rho\vec\nabla\cdot\vec{v}
\end{equation}

Examining the RHS of the latter equation, one finds that 
$\dot\rho=\dot\rho(\vec{x}, \vec{v})$. Since the referred equation is 
first-order total derivative we cannot neglect the fact that velocities and 
positions are desynchronized. One can write the following finite difference 
equation:
\begin{equation}\label{eq:104}
 \rho_{n+1/2} = \rho_{n} + \frac12\dot\rho_{n}\;\tau_{n+1/2} + 
\frac18\ddot\rho_{n}\;\tau_{n+1/2}^2
 +O(\tau_{n+1/2}^3)
\end{equation}
Similarly
\begin{equation}\label{eq:105}
 \rho_{n-1/2} = \rho_{n} - \frac12\dot\rho_{n}\;\tau_{n-1/2} + 
\frac18\ddot\rho_{n}\;\tau_{n-1/2}^2
 +O(\tau_{n-1/2}^3)
\end{equation}
Thus, subtracting (\ref{eq:105}) from (\ref{eq:104}), member to 
member, and rearranging, one has
\begin{equation}\label{eq:106}
 \rho_{n+1/2} = \rho_{n-1/2} + \dot\rho_{n}\;\overline{\tau}_{n} + 
\frac12\ddot\rho_{n}\;
 \overline{\tau}_{n}\delta{\tau}_{n}
 +O((\tau_{n+1/2}+\tau_{n-1/2})^3)
\end{equation}
On the other hand, one finds the first-order approximation:
\begin{equation}
 \dot\rho_{n+1/2} = \dot\rho_{n} + \frac12\ddot\rho_{n}\;\tau_{n+1/2} + 
O(\tau_{n+1/2}^2)
\end{equation}
and the backward solution for previous time level $n-1/2$,
\begin{equation}
 \dot\rho_{n-1/2} = \dot\rho_{n} - \frac12\ddot\rho_{n}\;\tau_{n-1/2} + 
O(\tau_{n-1/2}^2),
\end{equation}
so that the velocity-synchronized density rate is predicted as
\begin{equation}\label{eq:109}
 \dot\rho_n=\frac{\dot\rho_{n+1/2}+\dot\rho_{n-1/2}}2 + 
O((\tau_{n+1/2}+\tau_{n-1/2})^2).
\end{equation}
By analogy to equation~(\ref{eq:106}), one has
\begin{equation}
 \dot\rho_{n+1/2} = \dot\rho_{n-1/2} + \ddot\rho_{n}\;\overline{\tau}_{n} + 
 O((\tau_{n+1/2}+\tau_{n-1/2})^2)
\end{equation}
from which one has the first-order approximation for the second derivative of 
the density at time level $n$:
\begin{equation}\label{eq:111}
 \ddot\rho_{n}\;\overline{\tau}_{n} = \frac{\dot\rho_{n+1/2} - 
\dot\rho_{n-1/2}}2 +
 O((\tau_{n+1/2}+\tau_{n-1/2})^2)
\end{equation}
Gathering equations~(\ref{eq:106}), (\ref{eq:109}) and (\ref{eq:111}), one 
finally has the time step adaptive, second-order accuracy finite difference 
density evolution equation:
\begin{equation}\label{eq:112}
 \rho_{n+1/2} = \rho_{n-1/2} 
 + \dot{\overline\rho}_n\;\overline{\tau}_{n} 
 + \frac12\;{\delta\dot\rho_n}\;\delta{\tau}_{n}
 + O((\tau_{n+1/2}+\tau_{n-1/2})^3),
\end{equation}
where the following assignments were done:
\begin{equation}
 \dot{\overline\rho}_n=\frac{\dot\rho_{n+1/2}+\dot\rho_{n-1/2}}2
\end{equation}
and
\begin{equation}
 \delta\dot\rho_n=\frac{\dot\rho_{n+1/2} - \dot\rho_{n-1/2}}2.
\end{equation}
To recall, density rates $\dot\rho_{n+1/2}$ we re computed from 
equation~(\ref{eq:65a}).

One alternative approach although less efficient to update densities rather 
than integrating is recalling the self-regulating kNN procedure and then 
computing the new densities according to equation~(\ref{eq:61}). Since the 
previous self-regulating cluster has already been learnt, few adjustments are 
necessary for the new positions since the previous smoothing ellipsoids give 
tips on where and how the new ellipsoids should be. Such approach is just the 
spirit of machine learning.

The integration scheme for thermal energy conservation is analogous to the mass 
conservation's time difference scheme in equation~(\ref{eq:112}). Thus, one has 
the following time step adaptive, second-order accurate, finite-difference 
scheme for thermal energies:
\begin{equation}\label{eq:113}
 {u}_{n+1/2} = {u}_{n-1/2} 
 + \dot{\overline{u}}_n\;\overline{\tau}_{n} 
 + \frac12\;\delta\dot{u}_n\;\delta{\tau}_{n}
\end{equation}
where
\begin{equation}\label{eq:118}
 \dot{\overline{u}}_n=\frac{\dot{u}_{n+1/2}+\dot{u}_{n-1/2}}2
\end{equation}
and
\begin{equation}\label{eq:119}
 \delta\dot{u}_n=\frac{\dot{u}_{n+1/2} - \dot{u}_{n-1/2}}2.
\end{equation}
The specific thermal energy rates, $\dot{u}_{n+1/2}$ and 
$\dot{u}_{n-1/2}$, appearing in equations~(\ref{eq:118}) and (\ref{eq:119}), 
were computed from equation~(\ref{eq:78}). Obviously, both $\dot{u}_{n+1/2}$ 
and $\dot{u}_{n-1/2}$ depend on the position-synchronized velocities 
$\vec{v}_{n+1/2}$ and $\vec{v}_{n-1/2}$, as estimated in 
equation~(\ref{eq:102}).

It should be noted that mass (\ref{eq:112}) and energy (\ref{eq:113}) 
conservation equations require the respective density $\dot\rho_{n+1/2}$ and 
specific thermal energy $\dot{u}_{n+1/2}$ rates at the latest time level to be 
stored in the dataset $\mathcal{D}$. Latest individual time steps should also 
be 
stored to substitute $\tau_{n-1/2}$ in equations~(\ref{eq:97}) and 
(\ref{eq:98}) 
in the next time level. 
Thus, considering that the initial conditions come from an $N$-instance 
dataset, 
$N=|\mathcal{D}|$, each instance represents one particle having the following 
particle attributes: $(m, \vec{x}_{n+1/2}, \vec{v}_{n+1}, u_{n+1/2}, 
\dot{u}_{n+1/2}, \dot{\rho}_{n+1/2}, \tau_{n+1/2})$.
Densities are not stored in $\mathcal{D}$ since they must be estimated at the 
beginning of the integration scheme since the predicted density values would 
differ considerably from the value estimated in the equation~(\ref{eq:61}) as 
time passed within the root time-step $\Delta t$. To address, a variable time 
step, say $\tau_{n+1/2}$ is written as $2^l$th of $\Delta{t}$, namely, 
$\tau_{n+1/2}=2^{-l}\Delta{t}$, where $l$ is the time depth in the 
previously described binary hierarchical leapfrog 
\citep[see, e.g.,][]{MarinhoLepine:2000}.

\section{Application: Collapse and fragmentation of non-magnetic rotating gas 
spheres}\label{sec:10}

We have adopted equation~(\ref{eq:82c}) to reproduce the collapse of a 
protostar. According to the equation, the sphere is initially isothermal given 
the initial conditions under the adopted physical scales. As the denser parts 
is forming, the $\gamma$-index smoothly changes from 1.0 (isothermal) to 1.4 
(adiabatic). The isothermal gas reproduces the transparent phase of the 
collapse, and the adiabatic component reproduces the opaque and denser part of 
the collapsing cloud. It is only in the formation of adiabatic core that an 
adiabatic shock from the inside out is produced.

Several adiabatic (or almost adiabatic) lumps appeared during the accretion 
disc 
formation phase. These have an aspect that suggests the idea of protostars. 
However, many of these protostars are devoured by the central massive object. 
This occurs more quickly in the isotropic simulation than in the anisotropic. 
In the latter, the lumps survive for several periods of disk rotation. Also, in 
the initial phase of the disc, filamentary fragments appear in the anisotropic 
simulation. In the isotropic case, this is not evident. Additionally, the 
lumps occurrence were much more abundant in the anisotropic case. In 
general, isotropic simulation is better behaved, as if it were a blurred 
version of the anisotropic simulation. This last observation means that the 
anisotropic simulation is richer in high contrast details than the isotropic 
one.

\subsection{Code description}

The self-gravitating, anisotropic SPH code was developed in C and has the 
covariance octree and its associated methods as its central core. These methods 
are the procedure for constructing the tree and to perform the tree-descent for 
searching for self-regulating kNN clusters and searching for well-distant nodes 
for calculating the gravitational forces.

The tree-descent algorithm for kNN search is essentially the same as introduced 
by \cite{Marinho-Andreazza:2010}. Similarly, the tree-descent algorithm for 
gravity computation, regardless the spatial tessellation method, is the same as 
presented in previous works \citep{MarinhoLepine:2000,Marinho-et-al:2001} since 
both classical and covariance-based octrees have the same topology. The only 
difference here is the way as the tolerance criterion is implemented as shown 
in Section~\ref{sec:07}, involving the covariance-node geometry, besides the 
claimed accuracy increase in the present method in comparison to the cubical 
(or hyper-rectangular) tessellation of the classical octree.

\subsection{Physical scales}

The computational physical scales were conveniently chosen to perform a 
self-gravitating simulation of the non-magnetic collapse of a rotating sphere 
of molecular hydrogen gas. This roughly matches the dimensions of a dark 
molecular cloud to form something like the topology of an open star cluster. Of 
course, such a scenario is totally unrealistic. The dynamic effects of the 
magnetic field are being replaced by the very high rotation of the cloud as if 
it were initially a rigid body. This high rotation rate is not observed in 
molecular clouds, which, in general, are observed at small scales of rotation, 
which are compared to the shear effect of the disc rotation of our Galaxy 
\citep[e.g.][]{Phillips:1999}, and also in the galaxies M~33 
\citep[e.g.][]{Braine-et-al:2018} and M~51 \citep[e.g.][]{Braine-et-al:2019}.

Mass unit was scaled as $[m] = 222.1$ M$_\odot$, and length unit $[l] =$ 1 pc. 
Time unit was computed from the free-fall scale formula, namely $$[t] = 
\sqrt{\frac{[l]^3}{G [m]}},$$ where $G=6.674\times10^{-8}$ 
g$^{-1}$cm$^3$s$^{-2}$ is the universal gravitational constant. Thus, computing 
from the formula above and converting the result to Myr, one has $[t]=1.000$ 
Myr.

Velocity unit $[v]$ is derived from length and time units, $[v]=[l][t]^{-1}$, 
yielding $[v] = $ 0.9773 km s$^{-1}$. For the sake of curiosity, the speed 
of the light, in terms of computational units is given by $c = 306,753\; [v]$. 
Something is too wrong with the chosen time step when SPH particles 
representing volumes of molecular cloud reach values closer to $c$.

The computational unit for angular velocity was calculated as $[\Omega] = 
1.00\times10^{-6}$ rad yr$^{-1}\equiv 1.59\times10^{-7}$ revolution per year, 
which corresponds to 6.29 Myr per revolution, observing that the rotational 
component of velocity is given by $\vec{v}_\text{rot}=\vec\Omega\times\vec{r}$, 
with $\vec\Omega$ given, for instance, in radians per second, and $\vec{r}$ in 
centimeters to yield $\vec{v}_\text{rot}$ in centimeters per second.

Thermal specific energy unit is derived from the units given above 
$[u]=[m][l]^2[t]^{-2}[m]^{-1}=[l]^2[t]^{-2}=9.55\times10^{+9}
$ erg g$^{-1}$. From equation~(\ref{eq:82}), one can easily find the 
temperature for the specific thermal energy of $u=1\;[u]$ as $T=91.91$ Kelvin.

Density unit corresponds to $[\rho] = 1.503\times10^{-20}$ g cm$^{-3} =$ 
4,526 molecules cm$^{-3}$ for a 100\% molecular hydrogen.

\subsection{Simulation parameters}

The total number of SPH particles was $N=16,384$, and the preset number of 
nearest neighbors was $k=\frac12\sqrt{N}=64$, which is half the Poisson 
distribution error, estimated in counting the total number $N$ of particles 
within the spherical hull with the expected number density of 
$\nu=N/(\frac43\pi{R}^3)$.

The root time-step for the modified leapfrog was $\Delta{t}=1/1024= 
0.000488281$, whose maximum time depth was 12, which means that the deepest 
possible time-step was $\Delta{t}/2^{12}=1.192092285\times10^{-7}$. Along the 
entire experiment the maximum time depth reached was 8.

The aperture (tolerance) parameter for gravity estimation was 
$\theta=0.176777$, combined with the Aarseth softening length 
$\epsilon=0.0763842$.

The critical density in equation~(\ref{eq:82c}) was 
$\rho_c=22.1\;[\rho]\equiv100$ molecules cm$^{-3}$, which is a lower limit to 
molecular dark clouds \citep[e.g,][]{YORK200345}.

\subsection{Initial conditions}

The initial conditions reproduce the initial stages of a piece of dark molecular 
cloud collapse as if that piece were initially isothermal until reaching 
adiabatic clumps. This bimodal fluid model works to interrupt the collapse when 
the fluid changes from isothermal to adiabatic, favoring the appearance of 
protostar candidates.

The system is initially a spherically homogeneous particle distribution, 
randomly generated according to an expected constant density 
sphere of radius, $R=1\;[l] = $ 1 pc, total mass, $M=1\;[m] = $ 222.1 M$\odot$, 
having center of mass at $\langle{\vec{x}}\rangle=\vec0$ and mean velocity 
$\langle{\vec{v}}\rangle=\vec0$.

The sphere is dynamically cold, namely, with null velocity dispersion,
$\langle{|\vec{v}|^2}\rangle-|\langle{\vec{v}}\rangle|^2=0$ and rotates as a 
rigid body with angular velocity, $\vec\Omega=1\;\vec{\hat{z}}\;[\Omega]$. 
Thus, the particles velocity, namely rotational velocity field, was distributed 
as $\vec{v}(\vec{r})=\vec\Omega\times\vec{r}$.

Specific thermal energy was uniformly distributed as $u=0.1$, which 
corresponds to an initial uniform temperature of 9.191~K.

\subsection{Main results}

\begin{figure}[h!]
 \centering
 \includegraphics[width=9 cm,keepaspectratio=true]{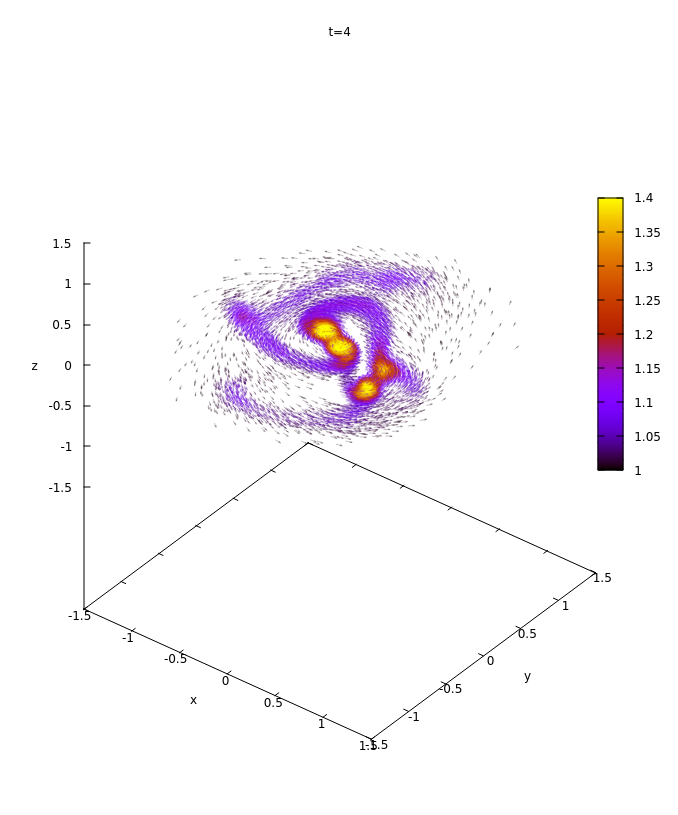}
 \caption{Isotropic simulation of the formation of a giant protostellar disc at 
t=4 (4096 time-steps). There are four protostars, but they are merging due to 
the resistance of the gas around them. The arrows denote the velocity field}
 \label{fig:02}
\end{figure}
\begin{figure}[h!]
 \centering
 \includegraphics[width=9 cm,keepaspectratio=true]{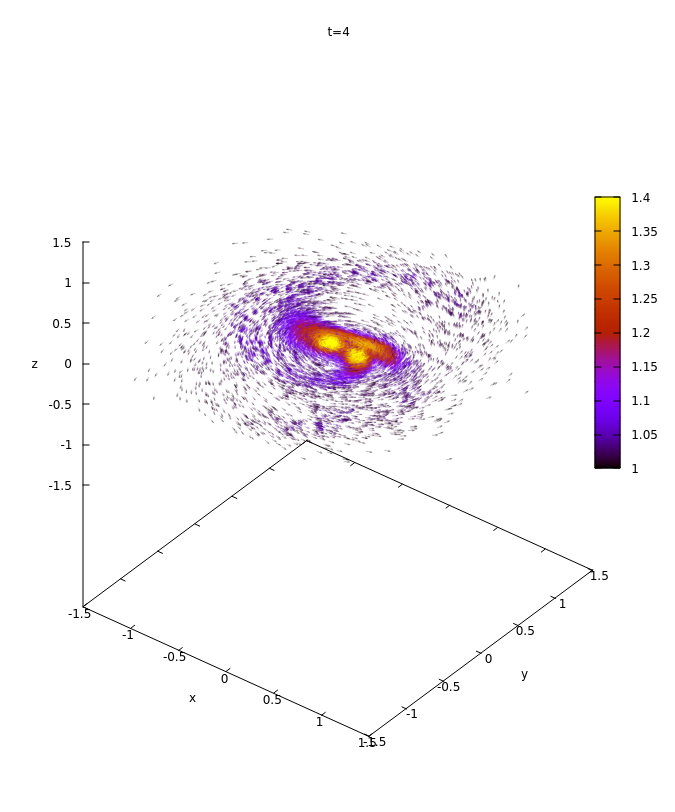}
 \caption{Anisotropic simulation of the formation of a giant protostellar disc 
at the same time of the previous figure, t=4 (4096 time-steps). There are three 
protostars, but they are merging due to the resistance of the gas around them, 
as in the isotropic case}
 \label{fig:03}
\end{figure}

Figures \ref{fig:02} and \ref{fig:03} show the disc formation at time $t=4$ 
(4096 time-steps) in both cases, isotropic and anisotropic respectively. At 
this time ($\sim$16 Myr) the system has collapsed to a giant protostellar disc 
with approximately 1 pc radius and $\sim0.01$ pc thickness. Comparing both 
figures, one can see clearly that the isotropic simulation presents a smooth 
distribution of particles representing the disc's isothermal component. On the 
other hand, anisotropic simulation is much more detailed and reveals the first 
stages of the isothermal disk fragmentation process. The four cores appearing 
in the isotropic simulation merges to form a solo core in the central part of 
the disc. The same happens to the three cores in the anisotropic case. The 
simulations could be more realistic if a gas-sweeping mechanism were introduced 
to mimic the effect of the protostellar wind. The color scale corresponds to 
the $\gamma$-index given in equation~(\ref{eq:82c}) where $\gamma=1$ (dark 
violet) stands for isothermal and $\gamma=1.4$ for adiabatic (yellow). The 
formed protostars appear in orange while the remnant debris are in light violet.

\begin{figure}[h!]
 \centering
 \includegraphics[width=9 cm,keepaspectratio=true]{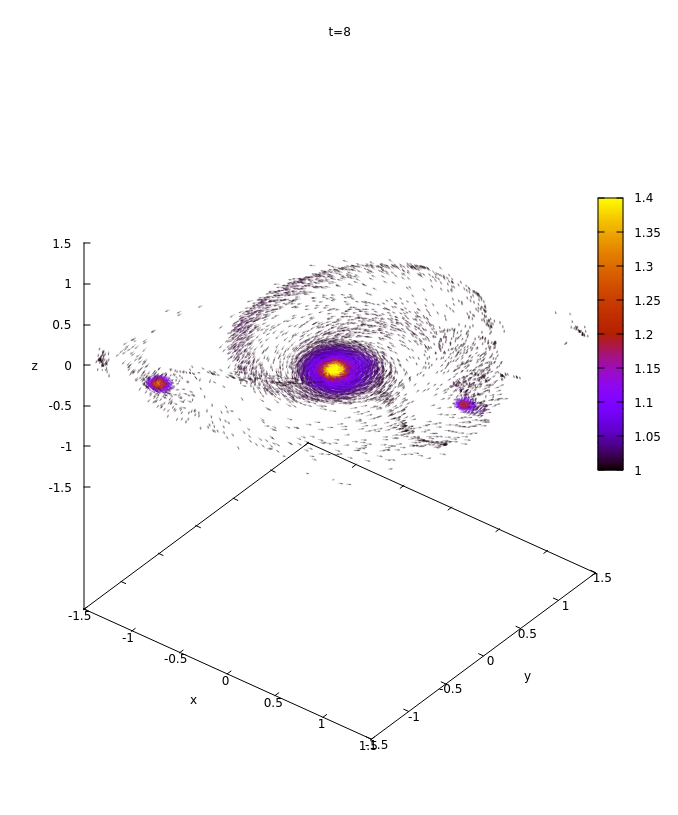}
 \caption{Isotropic simulation of the formation of a protostellar disc at t=8 
(8192 time-steps)}
 \label{fig:04}
\end{figure}
\begin{figure}[h!]
 \centering
 \includegraphics[width=9 cm,keepaspectratio=true]{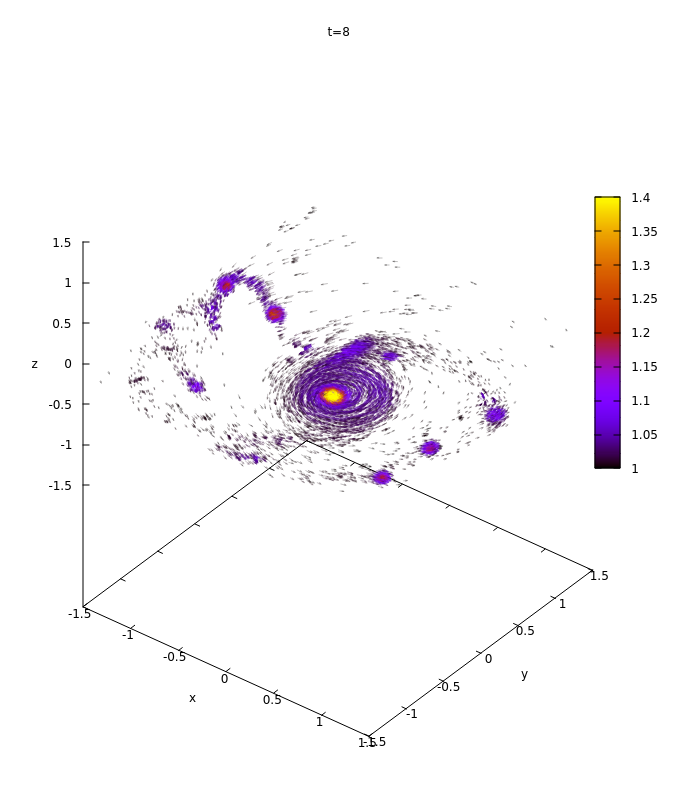}
 \caption{Anisotropic simulation of the formation of a protostellar disc at t=8 
(8192 time-steps)}
 \label{fig:05}
\end{figure}

Figures \ref{fig:04} and \ref{fig:05} reveal the disc fragmentation at time 
$t=8$ ($8192$ time-steps, $\sim$40 Myr). In the isotropic case, Fig. 
\ref{fig:04}, appear essentially two protostars, assuming the central object 
is the protostar, while in the anisotropic case, Fig. \ref{fig:05}, one can 
count more than 10 protostars. Such protostars are self-gravitating lumps 
of gas in an intermediate state between isothermal and adiabatic, say 
$\gamma=1.25$.

\begin{figure}[h!]
 \centering
 \includegraphics[width=9 cm,keepaspectratio=true]{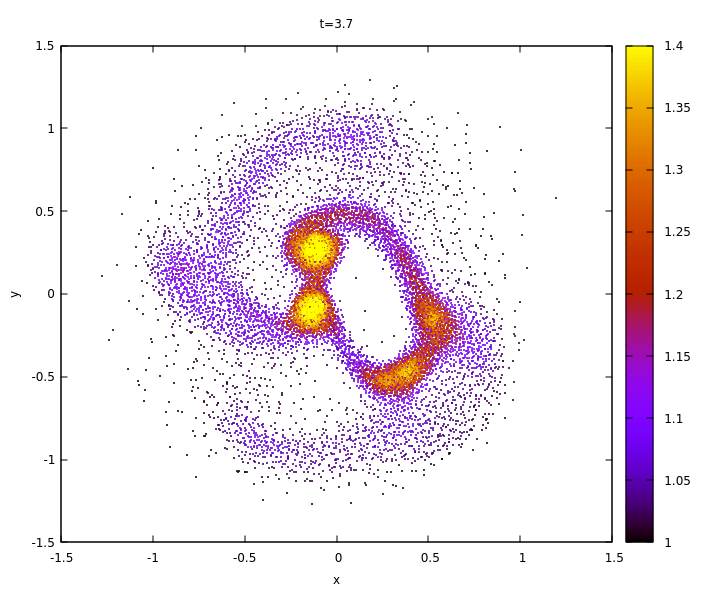}
 \caption{Isotropic case of the protostellar disc at t=3.7}
 \label{fig:06}
\end{figure}
\begin{figure}[h!]
 \centering
 \includegraphics[width=9 cm,keepaspectratio=true]{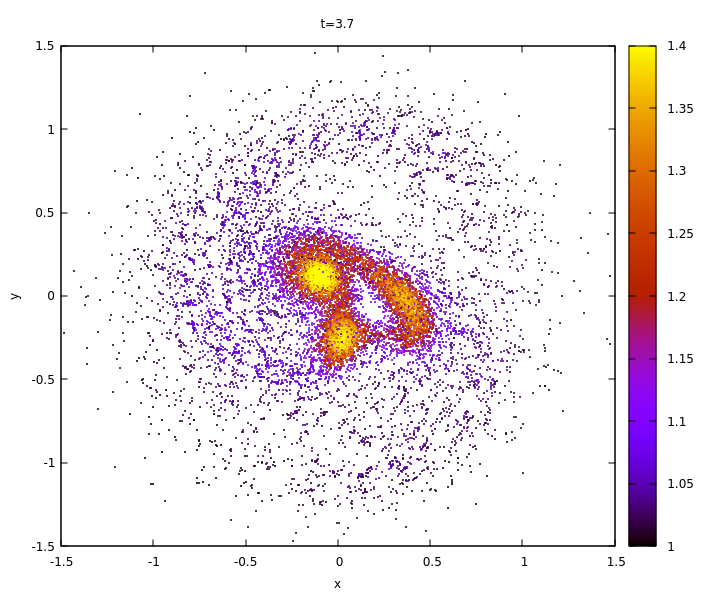}
 \caption{Anisotropic case of the protostellar disc at t=3.7 revealing 
filament-like fragments}
 \label{fig:07}
\end{figure}

Figures \ref{fig:06} and \ref{fig:07} show in more details the discrepant 
results among the isotropic and the anisotropic simulations, both at 
time $t=3.7$. In the latter, it is noticeable the large amount of 
filament-like fragments throughout the isothermal disc.


\section{Discussion and conclusion}\label{sec:11}

We have seen from the derivations made in Section~\ref{sec:05} that the 
anisotropic SPH equations have an invariant aspect concerning the classic SPH 
interpolation equations, changing only the anisotropic kernel derivatives and 
the resulting spatial resolution of the simulations shown in the 
Sec.~\ref{sec:10}.

When the anisotropic results were compared with the isotropic ones for the 
first 
time, the impression left was that the anisotropic simulation was quite 
imprecise, allowing particles' interpenetration. This apparent 
shock-overshooting did not seem intuitive since the artificial viscosity was 
designed to work effectively against the smoothing volume's flattening 
direction. For example, a smoothing volume in the shock layer is expected to be 
an oblate ellipsoid whose smaller semi-axis has the normal direction of the 
shock front, which is, of course, the direction in which the artificial 
viscosity is maximum. It was then that, after reviewing the Rayleigh-Taylor 
instability, it became clear that the evolution of shock in the anisotropic 
case 
was more realistic than in the isotropic case. The adoption of isotropic 
smoothing prevents or mitigates the effects of such two-fluid instability.

The simulation of a rotating self-gravitating gas's dissipative collapse 
evolved into a disc, thin on the inside and thick at the edges, similar to a 
protoplanetary disc. There was a relative loss of detail in the isotropic case. 
There was also much more formation of protostars in the anisotropic case due 
to the accretion shock's necessary resolution in the rotating collapse's 
critical phase. This result reinforces the idea that adopting anisotropic SPH 
simulations is fundamental, especially in astrophysics problems.

The use of smoothing volumes that adapt to the multivariate distribution of 
particles is equivalent to the MSVK method. In both methods, there was a favor 
in the formation of filaments since the smoothing ellipsoids tend to become 
prolate and aligned with the filaments. Likewise, the ongoing shock front is 
matched by the flattening of the ellipsoid in the shock direction. Thus, there 
is feedback in thinning both the filaments and the shock compared to isotropic 
simulations with spherical kernels.

The negative aspect of the proposed method is that it may be twice slower than 
the isotropic version, even when performing the non-gravitational simulation. 
This is due to the anisotropic artificial viscosity, which requires much shorter 
time steps according to Courant's stability criteria. This is easy to 
understand because the artificial viscosity reaches very high values in the thin 
shock layers, or filamentary structures, in the direction of the smallest 
principal component due to the flattening or elongation of the ellipsoidal 
support of the kernel function, increasing the magnitude of the kernel gradient 
in the direction of the minor principal component.

An additional observation for anyone interested in testing or giving 
contributions to the present code is that it is a prototype, not optimized for 
bolder purposes, which would otherwise require a complete review regarding 
optimization for high-performance computing as, for instance, proposed by 
\cite{10.1007/978-3-642-33078-0_6}. Moreover, the code is a console version, so 
it has no graphics user interface. However, creating a graphical interface is 
relatively simple with the nowadays features, especially on UNIX / Linux 
platforms, using, for example, the interface utilities provided by the GTK 
Project (\url{https://www.gtk.org/}). The code is available under e-mailed 
request to {pereira.marinho@unesp.br}.


%
\appendix 
\section{B-spline kernel}\label{app:A}

The 3D version of the B-spline kernel can be written as
\begin{equation}
 K_\text{3D}(\xi)=
 \frac6{\pi}
 \begin{cases}
    \frac43
    -8\xi^2
    +8\xi^3,             &0\le\xi\le\frac12\\
    \frac83(1-\xi)^3,   &\frac12\le\xi\le1\\
    0,                  &\xi\ge1
 \end{cases}
\end{equation}
so that the smoothing kernel is be computed by
\begin{equation}
 W_\tensor{H}(\vec{r})=\frac1{\text{det}\,\tensor{H}}
 K_\text{3D}(\lvert\tensor{H}^{-1}\vec{r}\rvert)
\end{equation}

Usually, in the isotropic case, the smoothing length $h$ is such that the 
smoothing kernel function vanishes at a distance greater than or equal to twice 
the length, which means $r/(2h)=1$ for a spherical compact support. Here, the 
B-spline curve has been adjusted to zero at a distance 
$\lvert\tensor{H}^{-1}\vec{r}\rvert=1$, which lies exactly on the surface of the 
ellipsoid hull under the replacement $r/(2h)\to\tensor{H}^{-1}\vec r$.

%

\end{document}